\documentclass[manuscript,revised]{geophysics}

\usepackage{color}
\usepackage{timet}
\usepackage{amsmath}
\usepackage[pdftex]{graphicx}

\usepackage{amsmath} 
\usepackage{amssymb,stmaryrd} 
\usepackage{amsfonts}
\usepackage{amscd}   

\usepackage{algorithm}
\usepackage{algorithmic}
\usepackage{setspace} 

\usepackage{gensymb}

\usepackage{epsfig}
\usepackage{multirow}
\usepackage{hhline}
\usepackage{multirow}

\def\RB{{\mathbb R}}

\begin{document}
\DeclareGraphicsExtensions{.pdf,.jpg,.mps,.png}



\title{A Data-Driven CO$_2$ Leakage Detection Using Seismic Data and Spatial-Temporal Densely Connected Convolutional Neural Networks}

\renewcommand{\thefootnote}{\fnsymbol{footnote}} 

\address{
\footnotemark[1]\textbf{Corresponding Author, ylin@lanl.gov}\\
\footnotemark[2]Earth and Environment Sciences Division\\ Los Alamos National Laboratory\\ Los Alamos, NM 87545 \\
\footnotemark[3]National Energy Technology Laboratory\\ United States Department of Energy\\ Pittsburgh, PA 15236\\
}
\author{Zheng Zhou\footnotemark[2], Youzuo Lin\footnotemark[1]\footnotemark[2], Zhongping Zhang\footnotemark[2], Yue Wu\footnotemark[2], Zan Wang\footnotemark[3], Robert Dilmore\footnotemark[3], and George Guthrie\footnotemark[2]}

\footer{Example}
\lefthead{Zhou  Et al.}

\label{firstpage}

\maketitle



\begin{abstract}
In carbon capture and sequestration (also known as carbon capture and storage, or CCS), developing effective monitoring methods is needed to detect and respond to CO$_2$ leakage.  CO$_2$ leakage detection methods rely on geophysical observations and monitoring sensor network. However, traditional methods usually require the development of site-specific physical models and expert interpretation, and the effectiveness of these methods can be limited to  different application locations, operational scenarios, and conditions. In this paper, we developed a novel data-driven leakage detection method based on densely connected convolutional neural networks. Our method is an end-to-end detection approach, that differs from conventional leakage monitoring methods by directly learning a mapping relationship between seismic data and the CO$_2$ leakage mass. To account for the spatial and temporal characteristics of seismic data, our novel networks architecture combines 1D and 2D convolutional neural networks. To overcome the computational expense of solving optimization problems, we apply a densely-connecting strategy in our network architecture that reduces the number of network parameters. Based on the features generated by our convolutional neural networks, we further incorporate a long short-term memory network to utilize time-sequential information, which further improves the detection accuracy. Finally, we employ our detection method to synthetic seismic datasets generated based on  flow simulations of a hypothetical CO$_2$ storage scenario with injection into a partially compartmentalized sandstone storage reservoir. To evaluate method performance, we conducted  multiple experiments including a random leakage test, a sequential test, and a robustness test. Numerical results show that our CO$_2$  leakage detection method successfully detects the leakage and accurately predicts the leakage mass, suggesting that it has the potential for application in monitoring of real CO$_2$ storage sites.

\end{abstract}
%

\section{Introduction}
The carbon capture and sequestration (CCS) technology collects the CO$_2$ from industrial sources such as thermal power plants and then injects compressed CO$_2$ into appropriate geologic formations underground. Storage operations must ensure that CO$_2$ is contained in the reservoir, even after injection stops (post-closure)~\citep{Yang-Probabilistic-2011}. Several monitoring technologies have been developed to detect CO$_2$ leakage at sequestration sites including observation of  seismic data, groundwater chemistry monitoring, near-surface measurements of soil CO$_2$ fluxes, analysis of carbon isotopes in soil gas, measurement of tracer compounds injected with the sequestered CO$_2$, and nearby atmospheric monitoring of CO$_2$ and tracer gases~\citep{Quantification-2011-Korre, Leuning-Atmospheric-2008, Benson-Monitoring-2007}. Among all these techniques, monitoring leakage through seismic data has been used extensively for plume mapping, quantification of the injected volume in the reservoir and early detection
of leakage~\citep{Geophysical-2011-Fabriol}.  Many related detection methods have sprung up in this field, such as obtaining the elastic parameters at different injection times through Gassmann fluid substitution to estimate the CO$_2$ sequestration status~\citep{Macquet-Feasibility-2017}. Most methods for interpreting seismic data rely on expert analysis, resulting in high costs and perhaps limiting detection at low volumes.

With rapid improvements in computational power and fast data storage, machine learning techniques have been effectively applied to problems from various domains. Deep learning, a technique with its foundation in artificial neural networks, is emerging in recent years as a powerful tool~\citep{Deep-2015-LeCun}. Among various deep learning methods, convolutional neural networks~(CNN) have achieved promising results in both detection and prediction tasks, such as speech recognition in 1D voice signal~\citep{Abdel-Convolutional-2014} and semantic segmentation in 2D image data~\citep{Jonathan-Fully-2015}. Long short-term memory (LSTM) networks, which are a type of the recurrent neural network~(RNN), and have achieved great success in the tasks associated with sequential data, such as natural language text compression~\citep{Martin-LSTM-2012} and time-series processing~\citep{Che-Recurrent-2018}. Compared with traditional RNNs and other sequence learning methods such as hidden Markov models~\citep{rabiner1989tutorial}, LSTM has shown its advantages in numerous applications through dealing with the exploding and vanishing gradient problem and keeping short-term ``memory'' for a long period of time ~\citep{Haşim-Long-2014}.

In this paper, we developed a novel end-to-end data-driven detection method, which directly learns the mapping relation from seismic data to CO$_2$ leakage mass as shown in Fig.~\ref{fig:LearningProcedure-Measurements-cut}. We designed our detection model based on conventional CNN and LSTM architectures. Seismic data comes with typical spatial and temporal characteristics.  Seismic traces from a single receiver is a typical 1D time series, while the 2D seismogram collected from multiple receivers can be treated as imagery, and there is spatial relevance between different traces. Therefore, instead of simply adapting the existing CNN architectures, we designed a novel network architecture by combining 1D and 2D CNN to account for both spatial- and temporal characteristics of seismic data.  Training a single CNN can be computationally expensive, and a combination of a couple of CNNs can be computationally prohibitive. To overcome the expensive computational cost, we built our network based on densely connected neural networks~\citep{huang2017densely, Wu-DeepDetect-2018} to reduce the network parameters. We call our specially designed neural networks ``spatial-temporal DenseNet'' (or ``ST-DenseNet'') in recognition that it is an extension of conventional DenseNet\citep{huang2017densely}.

CNN-based network structures are powerful in generating high-level features from the input signals at a single time stamp. The leakage of CO$_2$ is a continuous time-dependent process. Our monitoring technique takes into consideration sequential information, which is proved to be extremely useful for accurate monitoring and estimation of the CO$_2$ leakage mass. There have been a few traditional sequence models such as  Markov models~\citep{rabiner1989tutorial}, conditional random fields~\citep{lafferty2001conditional}, and Kalman filters ~\citep{brown1992introduction}. However, all these models have their limitations in learning long-range dependencies, and some of them require domain knowledge or feature engineering, offering less opportunity for serendipitous discovery~\citep{Dietterich2002Machine}. In contrast, neural networks are capable of learning high-level abstract representations automatically, and they can discover unforeseen features. In particular, LSTM~\citep{Long-Hochreiter-1997} models have achieved state-of-the-art results for many sequential problems of varying-length sequential data. We incorporate the LSTM structures to our models to account for the historical data. These structures can capture long-range dependencies and nonlinear dynamics, therefore further improve the accuracy of our monitoring method.

We implemented a sequence of computational experiments to evaluate the performance of our model including efficiency, accuracy and robustness  in various scenarios. We firstly showed that ST-DenseNet generates more effective features in comparison with other learning techniques. We further demonstrated that with the help of the LSTM structure, ST-DenseNet can overcome the discontinuity issue in sequential CO$_2$ leakage prediction results. To our best knowledge, there are no real seismic measurements available from an existing CO$_2$ storage site with small leakage happening. Therefore, all the numerical tests of our monitoring method were based on  synthetic seismic datasets generated using a model for a potential CO$_2$ storage site at Kimberlina, California~\citep{birkholzer2011sensitivity, Simulated-2017-Buscheck}. The Kimberlina site is in a partially compartmentalized sandstone basin. A model has been developed for the site ~\citep{birkholzer2011sensitivity} and this model has been used to simulate various leakage scenarios~\citep{Simulated-2017-Buscheck}. We also validated the robustness of our monitoring methods by utilizing a intra-site cross-location test (training and testing our model on the data from different locations) and noisy-data test. 

In the following sections, we first briefly describe the physics-driven detection methods,  data-driven detection methods, and their differences. We then introduce some basic architectures of the deep neural networks that are used in our methods. We further develop and discuss our novel  CO$_2$ leakage monitoring  method based on Densely Connected Network and LSTM. In the end, we apply our method to test problems using synthetic reflection seismic data and  discuss the results. Finally, concluding remarks are presented.

\section{Methodology}

\subsection{Physics-Driven and Data-Driven Detection Approaches}

\subsection{Physics-Driven Detection Approach}
In this work, we employ seismic data as the physical measurements and wave equations are the governing physics behind the problem. In the time-domain, the acoustic-wave equation is given by 
\begin{equation}
 \left [ \frac{1}{K(\mathbf{r})} \frac{\partial ^2}{\partial t ^2} 
        - \nabla  \cdot \left ( \frac{1}{\rho (\mathbf{r})}\,\, \nabla 
        \right ) \right ]
        p(\mathbf{r}, t) = s(\mathbf{r},\, t),
\label{eq:Forward}
\end{equation}
where $\rho (\mathbf{r})$ is the density at spatial location $\mathbf{r}$, $K(\mathbf{r})$ is the bulk modulus, $s(\mathbf{r},\, t)$ is the source term, $p(\mathbf{r}, t)$ is the pressure wavefield,  and $t$ represents time. 
The elastic-wave equation is written as
\begin{equation}
 \rho(\mathbf{r})\, \ddot{u}(r, t) - \nabla \cdot [C(\mathbf{r}) 
 : \nabla u(\mathbf{r}, t)] = s(\mathbf{r},\, t),
 \label{eq:ForwardElastic}
\end{equation}
where $C(\mathbf{r})$ is the elastic tensor, $u(\mathbf{r}, t)$ is the displacement wavefield, and operator ``:'' is the double dot product.

To simplify the notations of the wave equations in Eqs.~\eqref{eq:Forward} and \eqref{eq:ForwardElastic}, we rewrite them using forward operator form
\begin{equation}
 P = f(\mathbf{m}),
 \label{eq:ForwardLinearM}
\end{equation}
where  $P$ is the pressure wavefield for the acoustic case, $f$ is the forward operator, and $\mathbf{m}$ is the 
model parameter vector, including the density and velocity.  We use a time-domain staggered-grid finite-difference scheme to solve the wave equation~\citep{Tan-2014-Efficient}. This finite-difference scheme has  16$^\text{th}$-order accuracy in space and 4$^\text{th}$-order accuracy in time. 

To better understand the impact of the fluid injection to the change of the geophysical parameters such as velocity, we rewrite the P-wave and S-wave velocities as 
\begin{align}
v_p  = \sqrt{\frac{K + 4\mu/3}{\rho}},\\
v_s  = \sqrt{\frac{\mu}{\rho}},
\label{eq:leakageRelation}
\end{align}
where $v_p$ and $v_s$ are the P-wave and S-wave velocity, respectively; $K$ is  the bulk modulus of the rock; $\mu$ is the shear modulus of the rock.  When there is CO$_2$ leak happening in the subsurface storage formations, both the bulk modulus, $K$, and the density of the rock, $\rho$ will change. On the other hand,  the value of $\mu$ stays constant during the fluid saturation, according to \cite{Uber-1951-Gassmann} and  \cite{General-1941-Biot}. More details about the impact of CO$_2$ leakage to the rock physics can be found out in the work of \cite{wang2018modeling}. Therefore, subsurface velocity maps can be used to indirectly infer the leakage of the CO$_2$. Solving for velocity maps in  Eq.~(\ref{eq:ForwardLinearM}) is usually posed as a minimization problem
\begin{equation}
E(\mathbf{m}) = \underset{\mathbf{m}}{\operatorname{min}} \left\{ 
\delta d \right\}
= \underset{\mathbf{m}}{\operatorname{min}} \left \{ \left \| 
d - f(\mathbf{m})\right \| _2 ^2 \right \},
\label{eq:MisFit}
\end{equation}
where $d$ represents a recorded/field waveform dataset, $f(\mathbf{m})$ is the corresponding forward modeling result, $\delta 
d = \left \| d - f(\mathbf{m})\right \| _2 ^2$ is the data misfit, and $||\cdot ||_2$ stands for the $\text{L}_2$ norm.  Solving 
Eq.~(\ref{eq:MisFit}) yields a model $\mathbf{m}$ that minimizes the mean square difference between observed and synthetic waveforms.  
However, because of the limited data coverage, solving the inverse problem based on Eq.~(\ref{eq:MisFit}) is ill-posed. Moreover, because of the nonlinearity of the minimization in Eq.~\eqref{eq:MisFit}, the solution of the inverse problem may be trapped in a local minimum of the misfit function in Eq.~\eqref{eq:MisFit}. A regularization technique may alleviate the non-uniqueness and ill-posedness issues of the inverse problem.

In order to monitor the change of the velocity value due to the CO$_2$ leaks. Time-lapse seismic data can be collected to carry two independent inversions in (3) to obtain the changes $\delta \mathbf{m}$ in subsurface, that  is
\begin{equation}
\delta \mathbf{m} = f^{-1}(d_{\text{time2}}) -
                                     f^{-1}(d_{\text{time1}}),
 \label{eq:ConventionDiff}
\end{equation}
where $f^{-1}$ stands for inversion of waveform data, and 
$d_{\text{time1}}$ and $d_{\text{time2}}$ are datasets collected at 
two different times. 

\subsection{Data-Driven Detection Approach}

\begin{figure*}
	\centerline{
			\includegraphics[width=0.60\textwidth]{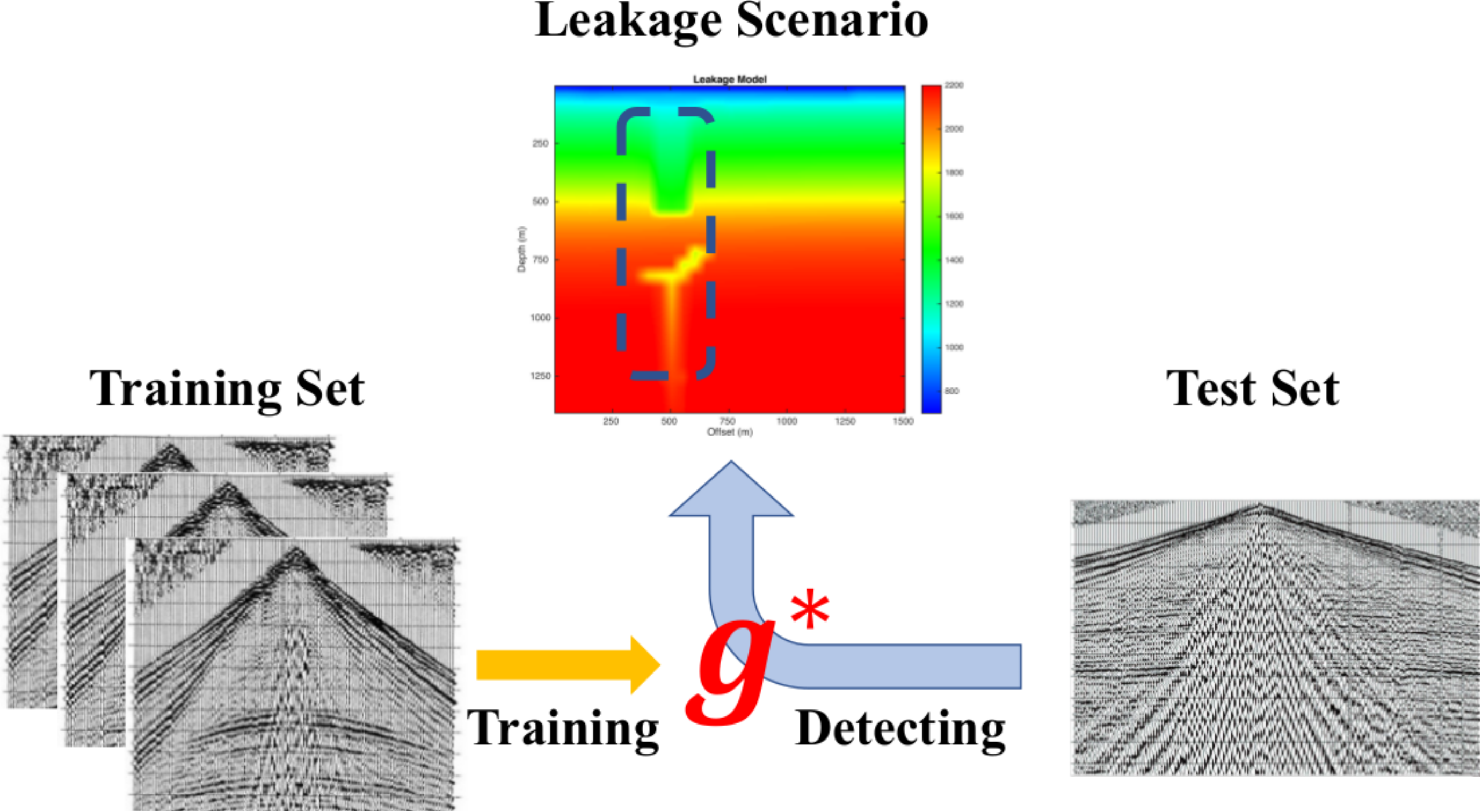}}
	\caption{The diagram  of the data-driven procedure to detect potential CO$_2$ leaks from seismic data.  Simulated seismic measurements  are utilized as training data sets, which are fed into the data-driven model. A mapping function, $g^\star (\cdot)$, 
is the outcome of the training algorithms. The function, $g^\star (\cdot)$, is the characterization function, which creates a link from the seismic measurements to the corresponding CO$_2$ leakage mass.}
	\label{fig:LearningProcedure-Measurements-cut}
\end{figure*}

In this paper, we adopt a data-driven approach, which means that we employ machine learning techniques directly to infer the geological changes from geophysical data. 
Overall, the idea of data-driven approach independent of applications can be illustrated as
\begin{equation*}
\displaystyle    \mathrm{\textbf{Physical Measurements}}  \xrightarrow[]{g^\star} \mathrm{\textbf{Labels}} .
\end{equation*}
Particularly in our problem, seismic data is utilized as the ``Physical Measurements'' and the leakage mass is the ``Labels.'' We build a machine learning model  using deep neural networks and train it using the synthetic and/or historical seismic measurements.
After training, one obtains a function, $g^\star$, which takes seismic measurements as input and returns a prediction of the leakage mass. Then for any unseen measurement, one can predict its corresponding leakage mass by $g^\star (\cdot)$. As for illustration, we provide our data-driven approach in Fig.~\ref{fig:LearningProcedure-Measurements-cut}. In the following sections, we provide more details on how we build a deep neural network to capture the CO$_2$ leakage mass from seismic data.

\subsection{Deep Neural Networks for Leakage Detection}
\label{section:Theory}

\subsection{Convolutional Neural Network}

Convolutional neural network~(CNN) is one of the most influential neural network structures in deep learning. LeNet  is known as the first kind of CNN~\citep{lecun1995learning}. In 2012, AlexNet won the ImageNet competition~\citep{krizhevsky2012imagenet}. The authors introduced fully connected layers and max-pooling layers to help AlexNet outperform all the other methods. After that, a sequence of different structures such as VGGNet~\citep{VGG}, ResNet~\citep{he2016deep}, GoogleNet~\citep{szegedy2017inception}, and DenseNet~\citep{huang2017densely} were introduced.  Regardless of the specific network architectures, all these different CNN networks consist of several common components including convolution layers, activation layers, batch normalization, and a loss function. Below, we provide some brief descriptions of these components. 

\subsubsection{Convolution Layer}
A convolution layer consists of various of filters that in traditional algorithms are hand-engineered. Compared with fully connected layers, a convolutional layer uses relatively little pre-processing because it only needs to learn the parameters of convolutional filters. The calculation requirements of CNN are also less than multi-layer perceptron~\citep{Geva1992constructive} because different positions of signals can share the convolution filters to extract features. We devised convolutional layers to extract the feature map of signals. The convolution can be defined for signals with arbitrary dimensions. The discrete convolution operation for 2-d data is defined as
\begin{equation}
X^{'}_{i, j} = \sum_{m}\sum_{n}K_{m, n} \cdot X_{(s-1)\times i+m, (s-1)\times j+n},
\end{equation}
where $X^{'}_{i, j}$ denotes the value of the $(i, j)$ location of the output signals, $X$ denotes the input signals, $K \in R^{m \times n}$ denotes the trainable kernel and $s$ denotes the stride between each sliding location of the kernel. 



\subsubsection{Activation Layer}
In neural networks, the activation layer is applied to introduce non-linearity, which helps the neural network increase the ability to learn complex mapping relationships. 
Nonlinear activation functions allow neural networks to compute nontrivial problems using only a small number of neurons. Some seminal activation functions such as Sigmoid and Relu have been widely used in deep learning~\citep{Liu2008One-Layer}. We provide some details on two different activation functions that are used in our method. 

\begin{enumerate}
\item Rectified Linear Unit (ReLU) Function: ReLU has been considered as one of the most useful activation functions to increase the sparsity and to alleviate the problem of gradient vanishing~\citep{nair2010rectified}. Its formula is defined as
\begin{equation}
\text{ReLU}(x) = \text{max}(0, x),
\end{equation}
where $x$ is the input vector which needs to be activated. In this paper, we used ReLU as the activation function in each convolution block.

\item Sigmoid Function: Sigmoid function is a mathematical function having a characteristic S-shaped curve. The sigmoid function is differentiable and monotonic. The sigmoid function can be a perfect choice to predict the probability as an output since its range is between 0 and 1. The sigmoid function is also widely used as an activation function in different deep learning models. Sigmoid function refers to the special case of the logistic function, which is defined by the formula
\begin{equation}
\text{Sigmoid}(x) = \frac{1}{1 + e^{-x}} =  \frac{e^{x}}{e^{x} + 1},
\end{equation}
where $x$ is the input vector that needs to be activated. The sigmoid function is applied to restrict the value range of high-level features before they are fitted into a regressor.

\end{enumerate}

%
 
\subsubsection{Batch Normalization}
Batch normalization is a technique for mini-batch gradient descent algorithms. It allows us to use much higher learning rates and to be less dependent with initialization. Batch normalization performs the following transformation for each activation $x_i$
\begin{equation}
B_{\gamma, \beta}(x_i) = \gamma \left(\frac{x_i - \mu}{\sqrt{\sigma^2 + \epsilon}} \right) + \beta,
\end{equation}
where $\gamma$ and $\beta$ are two trainable parameters; $i$ denotes the $i^{th}$ location of the feature map after a convolution layer; $\mu$ is calculated by averaging all values on the same feature map of $x_i$ given the mini-batch.

\subsubsection{Convolution Block}
The convolution block is usually a combination of several convolution layers (Conv), a batch normalization (BN), and an activation layer (ReLU). The convolution block can be summarized as a set of layers
\begin{equation}
x^{(l+1)} = \text{ReLU}(\text{BN}(\text{Conv}(x^{(l)}))),
\end{equation}
where $x^{(l)}$ and $x^{(l+1)}$ denote the input and output of the $l^{th}$ convolution block.

\subsubsection{Loss Function}

In this work, we utilized the mean squared error~(MSE) as the loss function to measure the distance between groundtruth and predicted values
\begin{align}
Loss_{MSE} = \frac{1}{n}\sum^{n}_{i=1}(Y_i-\hat{Y_i})^2,
\end{align}
where $n$ is the sample size, $\hat{Y} \in \RB^{1}$ is predicted leakage mass, and $Y\in \RB^{1}$ is the groundtruth.

\subsection{Residual Neural Network}
As for traditional deep neural networks, an important issue is the gradient vanishing problem, which means the gradient will be vanishingly small during the back-propagation procedure~\citep{Goodfellow-Deep-2016}. In particular, the gradient vanishing issue can effectively prevent the parameters from updating their values, which is due to the fact that the gradient decreases exponentially with the number of neural network layers~\citep{Goodfellow-Deep-2016}. To solve this issue, \cite{he2016deep} proposed residual neural network~(ResNet) framework which includes a shortcut in the convolutional neural network to avoid gradient vanishing.

Consider an input vector $x_0$ that is passed through a convolutional network. The network comprises $L$ layers, each of which implements a non-linear transformation $H_l(\cdot)$, where $l$ is the index of the layer. $H_l(\cdot)$ can be a composite function of operations such as batch normalization (BN), rectified linear units (ReLU), pooling, or convolution. We denote the output of the $l^{\mathrm{th}}$ layer as $x_l$. Traditional convolutional feed-forward networks~(such as VGG Net as shown in Fig.~\ref{fig:res_dense}(a)) connect the output of the $l^{\mathrm{th}}$ layer as input to the $(l+1)^{\mathrm{th}}$ layer, which gives rise to the following layer transition
\begin{equation}
x_l = H_l(x_{l-1}).
\label{eq:VGGNET}
\end{equation}
ResNet, As shown in Fig.~\ref{fig:res_dense}(b) , adds a skip-connection that bypasses the non-linear transformations with an identity function~\citep{he2016deep}
\begin{align}
x_l = H_l(x_{l-1})+x_{l-1}.
\label{eq:RESNET}
\end{align}

\subsection{Densely Connected Network}
\begin{figure}
\centering
\subfigure[]{\includegraphics[width=0.11\linewidth]{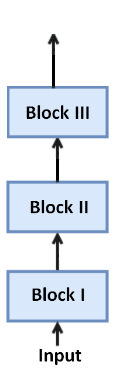}}
\hfil
\subfigure[]{\includegraphics[width=0.1\linewidth]{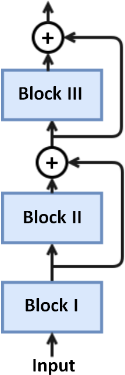}}
\hfil
\subfigure[]{\includegraphics[width=0.1\linewidth]{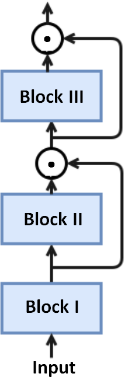}}
	\caption{A schematic illustration of the VGG Net (a), ResNet (b) and DenseNet (c), where ''$\rightarrow$'' represents information flow, ``$\bigoplus$'' represents feature addition operation and ``$\bigodot$'' stands for feature concatenation operation.}
	\label{fig:res_dense}
\end{figure}

The intuition behind densely connected networks~(DenseNet) is similar to that of ResNet~\citep{huang2017densely}. Both of them aim at reusing the convolution features from previous layers and reducing the number of trainable parameters. Different from ResNet, DenseNet concatenates the features from different layers instead of directly adding them together. DenseNet has many advantages. It alleviates the vanishing-gradient problem, reinforces the feature propagation, and substantially reduces the number of parameters~\citep{huang2017densely}. As shown in Figure~\ref{fig:res_dense}(c), a densely connected block is formulated as
\begin{align}
x_{l + 1} & = \mathcal{H}([x_{0}, x_{1}, ..., x_{l}]),\\
\mathcal{H}(x) & = W*(\sigma(B(x))),
\label{eq:denseblock}
\end{align}
where $W$ is the weight matrix, the operator of ``*'' denotes convolution, $B$ denotes batch normalization, $\sigma(x) = \text{max}(0, x)$ and $[x_{0}, x_{1}, ..., x_{l}]$ denotes the concatenation of all outputs of previous layers. 

\begin{figure*}
	\begin{center}
		\centering
		\includegraphics[width=0.75\columnwidth]{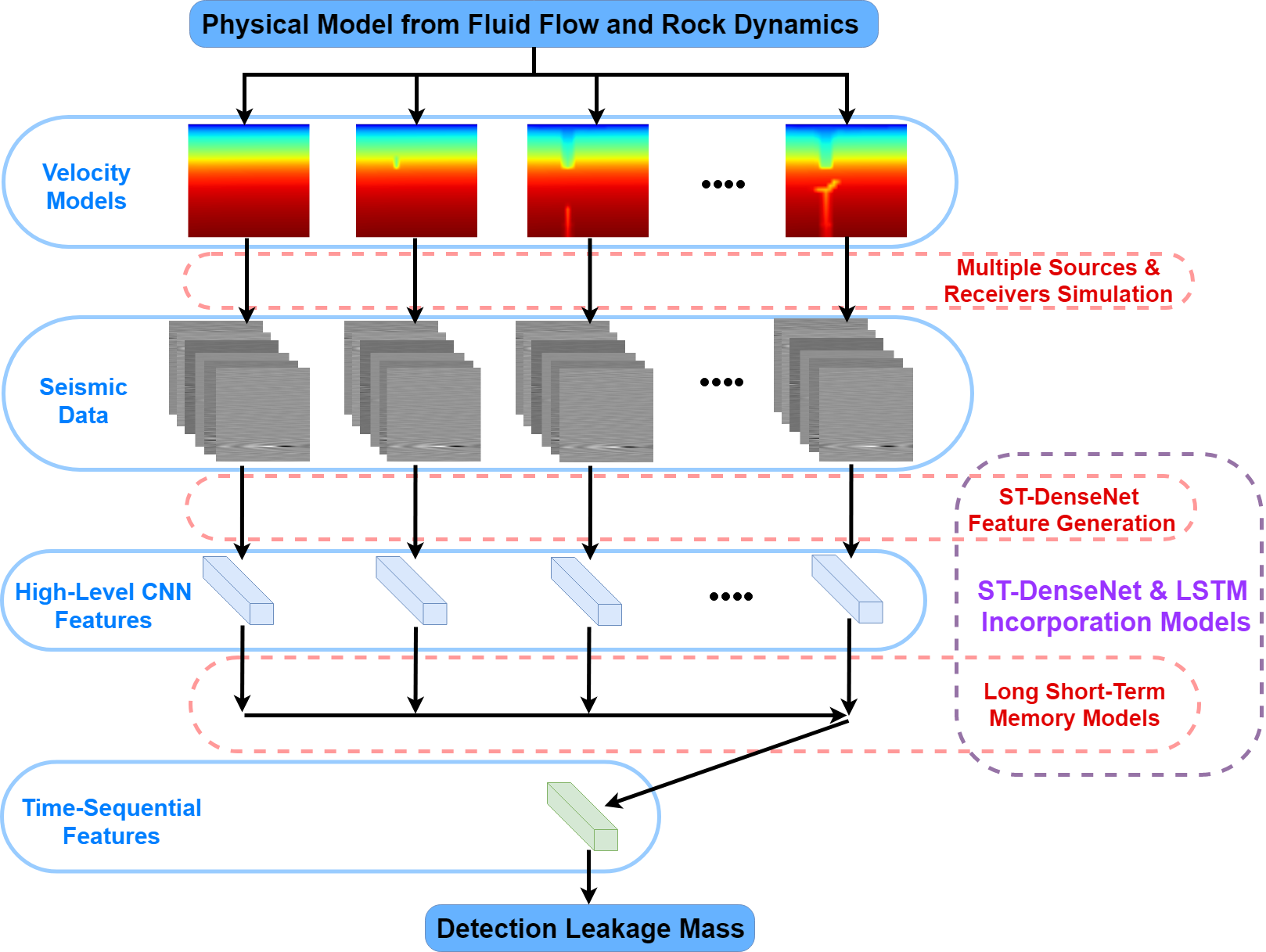}
	\end{center}
	\caption{The schematic illustration of our detection method. In the data generation stage, the velocity models are generated from the fluid and rock simulations, and the simulated seismic data are further obtained from these velocity models. Next stage is model training. Majority of seismic data along with actual CO$_2$ leakage mass are fit into ST-DenseNet model to generate high-level features to represent the original information in seismic data. Then long short-term memory networks can merge these high-level features into time-sequential features, and the regressor will be used to detect the CO$_2$ leakage mass based on seismic data.}
    \label{fig:ST-DensNet}
\end{figure*}

\section{Spatial-Temporal DenseNet}
\label{sec:STDenseNet}

The seismic trace from a single receiver is a typical 1D time series, while the 2D seismogram collected from multiple receivers can be treated as imagery, and there is spatial relevance between different traces. In order to account for both the spatial and temporal characteristics, we developed a new network structure, called ``spatial-temporal DenseNet (ST-DenseNet)'', by incorporating different CNN architectures.  We provide a schematic illustration of our  ST-DenseNet  in Fig.~\ref{fig:ST-DensNet}. Our ST-DenseNet model generates effective features by taking into consideration the spatial-temporal characteristics of seismic data. With the feature generated, we can further improve our model by incorporating a long short-term memory (LSTM) network. We will elaborate on the architecture of ST-DenseNet and how to incorporate LSTM. 

\subsection{ST-DenseNet: Feature Generation Using 1D/2D CNNs}

Our ST-DenseNet is built upon DenseNet. We provide its specific network architecture in Tabel~\ref{table:densenet_architect}. The major difference between conventional DenseNet and ST-DenseNet is that our model starts with applying convolution layers to 1D time series, followed by employing convolutions on the 2D seismogram. There are a couple of benefits of using ST-DenseNet. Firstly, applying 1D convolution layers in the time domain cannot only reduce the size of model parameters but also facilitates the learning of important temporal features. As shown in Table~\ref{table:densenet_architect}, ST-DenseNet results in a reductionality of $10^4$ for the example simulated seismic data we used in this study. Secondly, applying 2D convolution layers fuses high-level spatial and temporal features together. All these benefits turn out to be critical in improving detection accuracy and reducing training costs.

\begin {table}
\centering
\begin{tabular}{ |c|c|c| }
\hline
Stage & Layers & Dim.\\
\hline
Input & - & 6000 $\times$ 100 $\times$ 6 \\
\hline
Conv1D-1 & conv(7$\times$1), 32, /(4$\times$1) & 1499 $\times$ 100$\times$32 \\
\hline
Conv1D-2 & conv(5$\times$1), 32, /(3$\times$1) & 499 $\times$ 100 $\times$ 32 \\
\hline
Pool1D & max-pool(2$\times$1), /(2$\times$1) & 249 $\times$ 100$\times$32 \\
\hline
Conv1D-3 & conv(3$\times$1), 32, /(2$\times$1)) & 124 $\times$ 100 $\times$ 32 \\
\hline
Dense2D-1 & [conv(3$\times$3), 64] $\times$ 3 & 124 $\times$ 100 $\times$ 224 \\
\hline
\multirow{2}{*}{Transition} & conv(1$\times$1), 64 & 124 $\times$ 100 $\times$ 64 \\
\hhline{~--} & max-pool(2$\times$2), /(2$\times$2) & 62 $\times$ 50 $\times$ 64 \\
\hline
Dense2D-2 & [conv(3$\times$3), 128] $\times$ 3 & 62 $\times$ 50 $\times$ 448 \\
\hline
\multirow{2}{*}{Transition} & conv(1$\times$1), 128 & 62 $\times$ 50 $\times$ 128 \\
\hhline{~--} & max-pool(2$\times$2), /(2$\times$2) & 31 $\times$ 25 $\times$ 128 \\
\hline
Dense2D-3 & [conv(3$\times$3), 256] $\times$ 3 & 31 $\times$ 25 $\times$ 896 \\
\hline
\multirow{2}{*}{Transition} & conv(1$\times$1), 256 & 31 $\times$ 25 $\times$ 256 \\
\hhline{~--} & max-pool(2$\times$2), /(2$\times$2) & 15 $\times$ 12 $\times$ 256 \\
\hline
Dense2D-4 & [conv(3$\times$3), 512] $\times$ 3 & 15 $\times$ 12 $\times$ 1792 \\
\hline
\multirow{2}{*}{Transition} & conv(1$\times$1), 512 & 15 $\times$ 12 $\times$ 512 \\
\hhline{~--} & max-pool(2$\times$2), /(2$\times$2) & 7 $\times$ 6 $\times$ 512 \\
\hline
Dense2D-5 & [conv(3$\times$3), 1024] $\times$ 3 & 7 $\times$ 6 $\times$ 3584 \\
\hline
\multirow{2}{*}{Transition} & conv(1$\times$1), 1024 & 7 $\times$ 6 $\times$ 1024 \\
\hhline{~--} & max-pool(2$\times$2), /(2$\times$2) & 3 $\times$ 3 $\times$ 1024 \\
\hline
Flatten Layer & conv(1$\times$1), 512 & 1 $\times$ 1 $\times$ 512\\
\hline
\multicolumn{3}{|c|}{1-d fully connected, mean squared error loss} \\
\hline
\end{tabular}
\caption{The feature generation stage of ST-DenseNet. This model is designed for inputs with 6,000$\times$100$\times$6, which are the simulated seismic data. ''Conv(7$\times$1), 32, /(4$\times$1)'' denotes using 32 (7$\times$1) convolution kernels with a stride of (4$\times$1).}
\label{table:densenet_architect}
\end {table}

\subsection{Incorporation of Long Short-Term Memory Network}

The leakage of CO$_{2}$ is a time-dependent accumulated procedure. To further account for the previous information for the learning task at a present time point, we utilize the recurrent neural networks (RNNs) structure to take the recent sequential information into consideration.  There are two popular RNN frameworks proposed: gated recurrent units (GRU)~\citep{chung2014empirical} and long short-term memory network (LSTM)~\citep{Long-Hochreiter-1997}. In this work, we select LSTM due to its supreme performance.

\subsubsection{Long Short Term Memory Network}
\begin{figure}
	\begin{center}
		\centering
		\includegraphics[width=1\columnwidth]{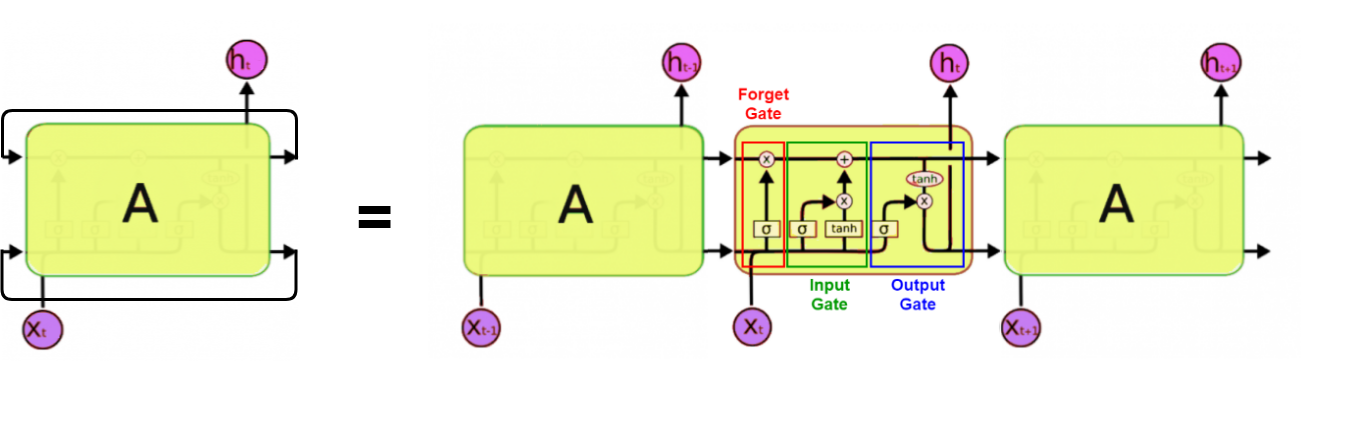}
	\end{center}
	\caption{An illustration of unrolling the long short-term memory networks, where $\sigma$ denotes the sigmoid activation function, $+$ represents the point-wise plus operation~\citep{Olah2015understanding}.}
    \label{fig:LSTM}
\end{figure}

Long short-term memory network is a special case of RNN models. It has a chain structure similar to the basic RNNs, but the repeating modules are organized in a completely different structure. Distinguished from the other traditional RNN models, the basic unit of the LSTM hidden layer is a memory block ~\citep{Zen2015Unidirectional}. The memory block (the outermost rectangle shown in the Figure \ref{fig:LSTM}) consists of memory cells that can memorize the temporary state, and a group of self-adaptive gating units to control information flow inside the memory block. Two gates, known as the input gate and the output gate, are added respectively to control the input and output in the block. The most important part of a memory cell is a recurrently connected linear unit, i.e., constant error carousel~(CEC)~\citep{Gers2000Learning}, whose activated value can represent the state of cells. Because of the existence of this structure, multiplicative gates can learn to open and close, and thus the LSTM networks can solve the problem of vanishing error and lacking long-term dependence by remaining the network error in a constant range. To prevent the internal cell values growing drastically when processing continuous sequential series without being previously segmented, a forget gate is introduced into the memory block. Similar to a human tendency to forget useless information, the forge gate allows the memory block to update by itself once the memorized information is out of date, and replaces the cell state by 
multiplying the activated value from forget gate. The entire computation can be summarized by a series of equations as follows
\begin{align}
f_t & =  \sigma (W_f \cdot [h_{t-1}, x_t] + b_f),\\
i_t & =  \sigma (W_i \cdot [h_{t-1}, x_t] + b_i),\\
\tilde{C_t} & =  \tanh (W_C \cdot [h_{t-1}, x_t] + b_C),\\
C_t & =  f_t * C_{t-1} + i_t * \tilde{C_t},\\
o_t & =  \sigma (W_0 \cdot [h_{t-1}, x_t] + b_0),\\
h_t & =  o_t * \tanh(C_t),
\end{align}
where $\sigma$ represents the sigmoid activation function, and $\rm tanh$ represents the hyperbolic tangent function. $x_t$ and $h_t$ represent the input and output vector at time step $t$. $W_f$, $W_i$ and $W_0$ are trainable weight matrices. $b_f$, $b_i$ and $b_0$ are corresponding bias terms. $C_t$ and $\tilde{C_t}$ are memory vectors, which can be used to remember long-term information.

\section{Numerical Experiments}
\label{sec:Results}

In this section, we provide three different numerical tests using synthetic seismic data generated based on flow simulations for a model CO$_2$  storage site~\citep{Simulated-2017-Buscheck, yang2018effectiveness}. We designed three tests to validate the performance of our CO$_2$ leakage detection method for different monitoring scenarios. In particular, we designed Test~1 to mimic the monitoring situation when a single seismic dataset is available. We designed Test~2 to mimic the monitoring situation when a sequence of seismic datasets is available over time. We further validated the performance of our detection method when data are noisy or training and testing data are not acquired from the same site.

\begin{figure}
	\begin{center}
		\centering
		\includegraphics[width=0.6\columnwidth]{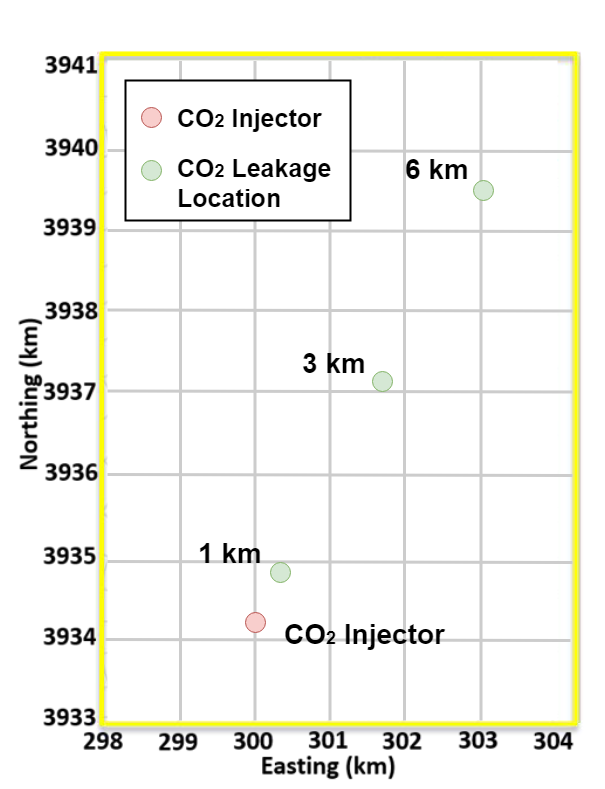}
	\end{center}
	\caption{Aquifer impact data is generated at three distances (1, 3, and 6 km) from the CO$_2$  injector using the hydrostratigraphy \citep{Simulated-2017-Buscheck}.}
    \label{fig:CO2-injector}
\end{figure}

\subsection{Dataset}
To our best knowledge, there are no real seismic measurements available from an existing CO$_2$ storage site with small leakage happening. Therefore, in order to evaluate the effectiveness of our developed approach, we test and validate our model for CO$_2$ leakage mass detection task on simulated reflection seismic datasets. The seismic velocity models were generated based on flow simulations for hypothetical brine and CO$_2$ leakage along damaged wellbores at a model CO$_2$ storage site near Kimberlina in the southern San Joaquin Basin, California. The legacy wells in the flow simulations are located at 1, 3, and 6 km away from the CO$_2$ injector. Most CO$_2$ has reached the very top layer of the model domain and CO$_2$ is in the gas phase in these wellbore leakage simulations. The set of flow simulations is specially developed for evaluating the effectiveness of CO$_2$ leakage monitoring methods. At each time step in the flow simulation outputs, we constructed seismic velocity models using the Gassmann’s equations ~\citep{gassmann1951verteljahrsschrift}. The porosity of the rock is assumed to be 0.35 in the Kimberlina wellbore leakage simulations. The physical properties of the effective pore fluid (a mixture of CO$_2$ gas and brine) depend on the temperature, pore pressure, salinity of brine and CO$_2$/brine saturation conditions. The temperature is kept constant at 40$^{\circ}$C in the wellbore leakage simulations. The pore pressure, the salinity of brine, CO$_2$ saturation and density of the CO$_2$ gas phase are obtained from the leakage simulation outputs. We provide in Table~\ref{table:physics-parameter} below  some representative values for the fluid and rock properties to generate simulations in our numerical tests. The workflow of constructing seismic velocity models based on flow simulations has been described in the work by~\citet{Modeling-2018-Wang}.  
\begin{table}
\centering
\begin{tabular}{ |c|c| }
\hline
Parameter Name & Parameter Value\\ 
\hline
\hline
Salinity of Brine&  0.001 \\
\hline
Density of Brine& 1 g/cm3   \\
\hline
Density of $\mathrm{CO}_2$ Gas& 0.075 g/cm3   \\
\hline
$\mathrm{CO}_2$ Fraction in the Pores & (0, 0.56) \\
\hline
Porosity of the Rock& 0.35 \\
\hline
Bulk Density of the Rock& 1.88 g/cm3\\
\hline
Bulk Modulus of the Rock& 1.18 GPa\\
\hline
Bulk Modulus of Mineral Grains& 31.1 GPa\\
\hline
Bulk Modulus of Dry Rock Frame& 1.15 GPa\\
\hline
\end{tabular}
\caption{The representative values for the fluid and rock properties used in our numerical tests to generate simulations.}
\label{table:physics-parameter}
\end{table}

To generate the seismic data, we assumed a total of 3 sources and 100 receivers evenly distributed along the top boundary of the model. The source interval was $500$~m, and the receiver interval was $15$~m. We used a Ricker wavelet with a center frequency of $50$~Hz as the source time function and a staggered-grid finite-difference scheme with a perfectly matched layered absorbing boundary condition to generate synthetic seismic reflection data~\citep{Tan-2014-Efficient, Zhang-2010-Unsplit}. The synthetic trace at each receiver consisted of a collection of 2-component time-series data, each of length $6,000$. With the seismic data available, the dimension of input data to our networks was $6,000 \times 100 \times 6$ (as illustrated in Fig.~\ref{fig:ST-DensNet}), where each of the 3 dimensions corresponds to time steps, receiver number, and component/source, respectively. 

The  CO$_2$ leakage mass was varied from $0$ to $10^{10}$~kilogram~(kg). We scaled all the mass data by a Log function as
\begin{equation}
\tilde{Y} = \log_{10}(Y+1), 
\end{equation}
where $Y$ represents the original value of CO$_2$ leakage mass, and $\tilde{Y}$ stands for scaled leakage mass. A value of 1 was added to each $Y$ to avoid taking a $\log(0)$.
All computations were carried on a computer with an Intel Xeon E5-2650 core running at 2.3~GHz, and Tesla K40c GPU with 875~MHz boost clock.

\subsection{Tests on Random Leakage Monitoring}

\label{Experiment 1}
\begin{table}
\centering
\begin{tabular}{ |c|c|c|c|c| }
\hline
& $\pm10\%$ Acc & $\pm5\%$Acc & $\pm3\%$Acc & Parameters \\ 
\hline
Kernel SVR & 0.263 & fail & fail & 36K \\
\hline
VGG-based CNN & 0.719 & 0.673 & 0.651 & 29M \\
\hline
ResNet-based CNN & \textbf{0.938} & 0.885 & 0.852 & 17M \\
\hline
ST-DenseNet & 0.936 & \textbf{0.912} & \textbf{0.891} & 9M \\
\hline
\end{tabular}
\caption{The CO$_2$ leakage mass prediction results given by kernel support vector regression (kernel SVR), VGG-based CNN, ResNet-based CNN and ST-DenseNet. ``fail'' indicates the accuracy is extremely low. The results indicate that ST-DenseNet outperforms both the classical regression model and other CNN-based models.}
\label{table:results1}
\end{table}

We tested the performance of ST-DenseNet provided with a single seismic dataset. The purpose of this test was to mimic the monitoring situation when the single seismic dataset is available. Specifically, we randomly select 2,400 groups for training, 300 groups for validating, and 227 groups for testing. We compared our method to different regression models, including 1. support vector regression with radial basis function kernel~(Kernel SVR); 2. VGG-based~CNN; and 3. ResNet-based CNN. We used accuracy with different error rates to evaluate the regression results. In particular, ``$\pm r\%$ Accuracy'' represents the tolerance of error rate, and this metric can be mathematically defined as
\begin{equation}
1-r\% \leqslant \frac{Predicted \quad Value}{Groundtruth \quad Value} \leqslant 1+r\%.
\end{equation}
In other words, if the predicted leakage mass is in the range from (100-r)\% to (100+r)\% of the actual leakage mass, the prediction result is considered as accurate. Otherwise, the prediction fails. In our test, we selected three different values of $r=3, 5$ and $10$ as evaluation metrics. The number of trainable parameters was also reported to reflect the computational complexity of different models.

The results from several different methods are summarized in Table~\ref{table:results1}. The kernel SVM had an extremely low accuracy, so it is listed as ``fail'' in the table. This result suggests that advanced feature extraction techniques are required in advance to apply SVR or other classical regression models. By comparing to the VGG-based CNN or ResNet-based CNN, ST-DenseNet yields higher accuracy. The only exception was the testing scenario of ``$\pm10\%$ Accuracy'', where our method still produces comparable results to those obtained by using ResNet-based CNN. The number of trainable parameters can be used as an indication of the computation complexity. In Table~\ref{table:results1}, we observe that ST-DenseNet yielded the smallest number of model parameters among all the CNN-based methods. So, using single seismic dataset, ST-DenseNet cannot only produce the most accurate CO$_2$ leakage mass detection but also requires the least amount of trainable parameters.  

\begin{figure*}
\centerline{
\subfigure[]{\includegraphics[width=0.50\linewidth]{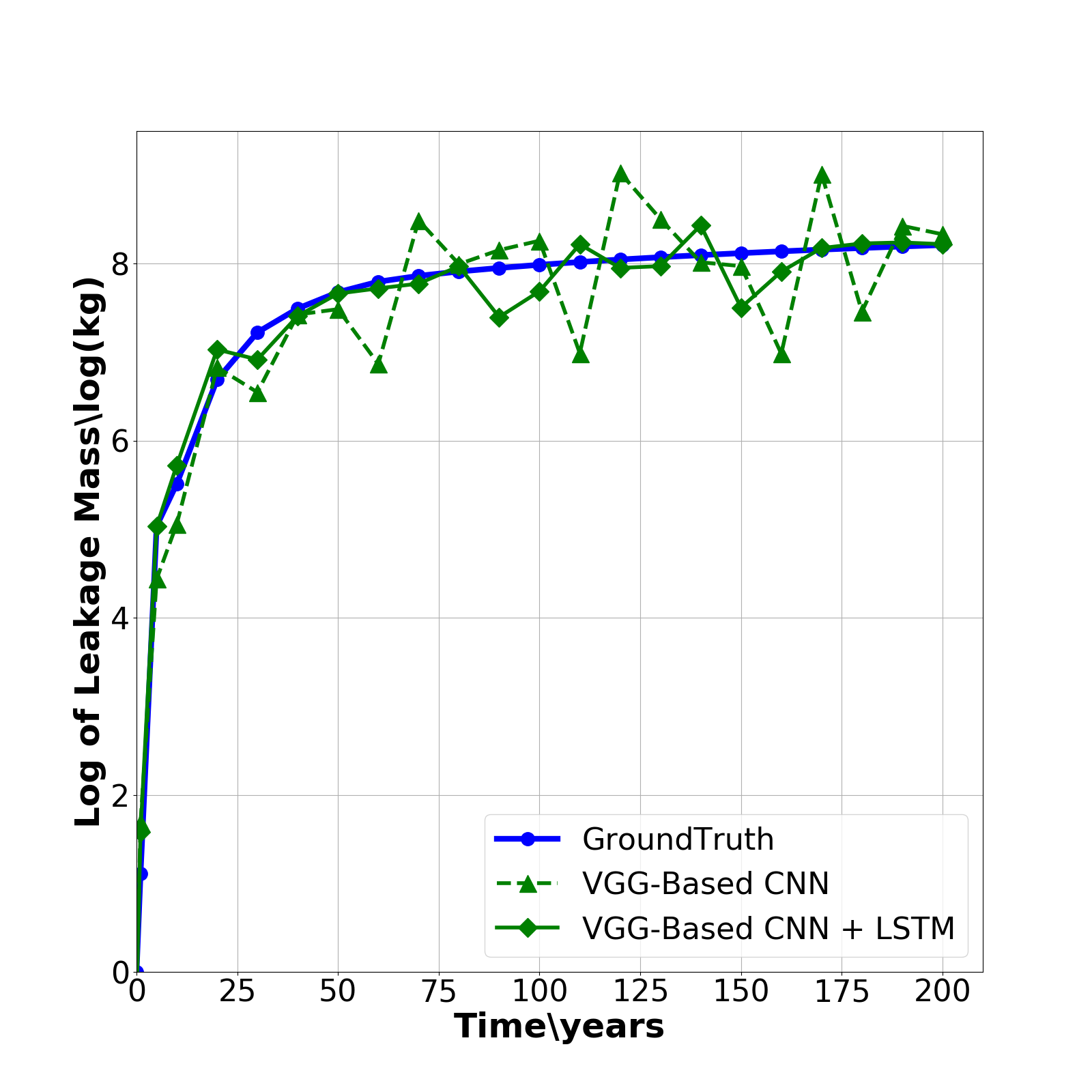}}
\subfigure[]{\includegraphics[width=0.50\linewidth]{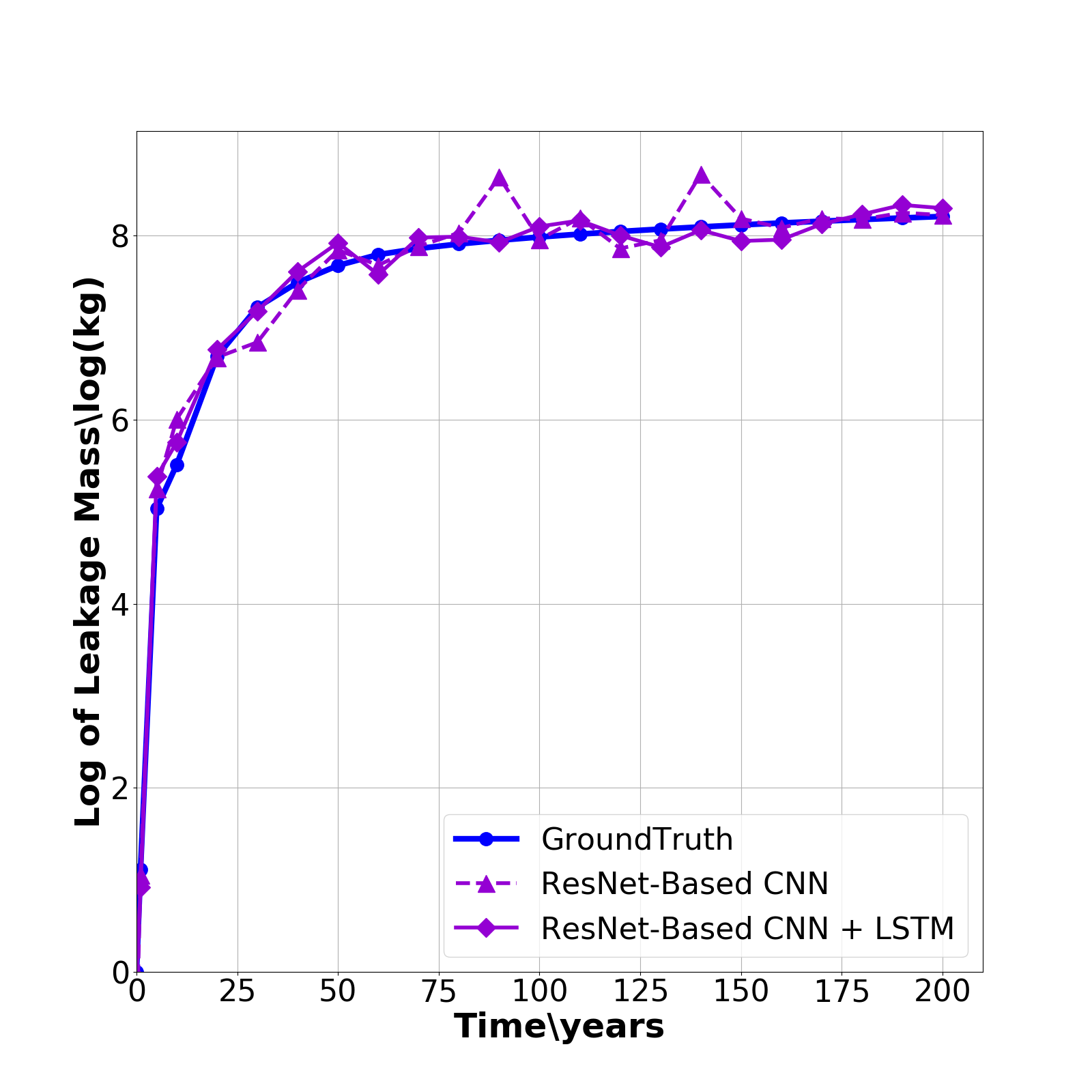}}}
\centerline{
\subfigure[]
{\includegraphics[width=0.50\linewidth]{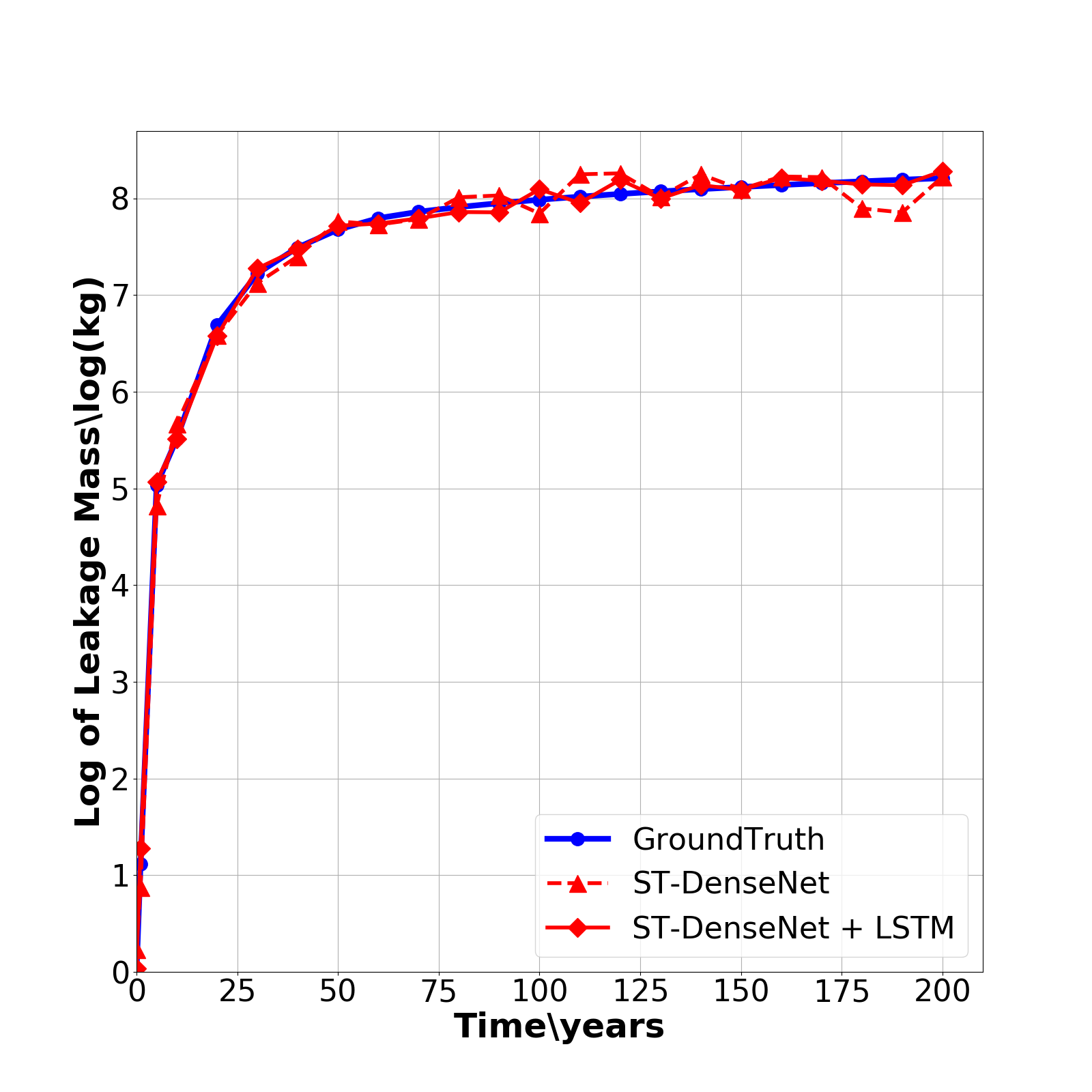}}
}
\caption{Illustration of three methods to detect CO$_2$ leakage mass at location 1~km away from injector. We provide detection results using three network architectures including VGGNet~(a), ResNet~(b), and ST-DenseNet~(c). Within each sub-figure, we also provide detection results using network with LSTM (in ``$-\diamond-$'') and without LSTM (in ``$-\triangle-$''). The ground-truth is plotted in blue.We observe that (1) the incorporation of LSTM improves the detection accuracy for all three CNN architectures including ST-DenseNet; (2) ST-DenseNet yields the most accurate detection results of all three methods.}
\label{fig:exp_detection_mass1}
\end{figure*}

\begin{figure}
\centering
\centerline{
\subfigure[]{\includegraphics[width=0.5\linewidth]{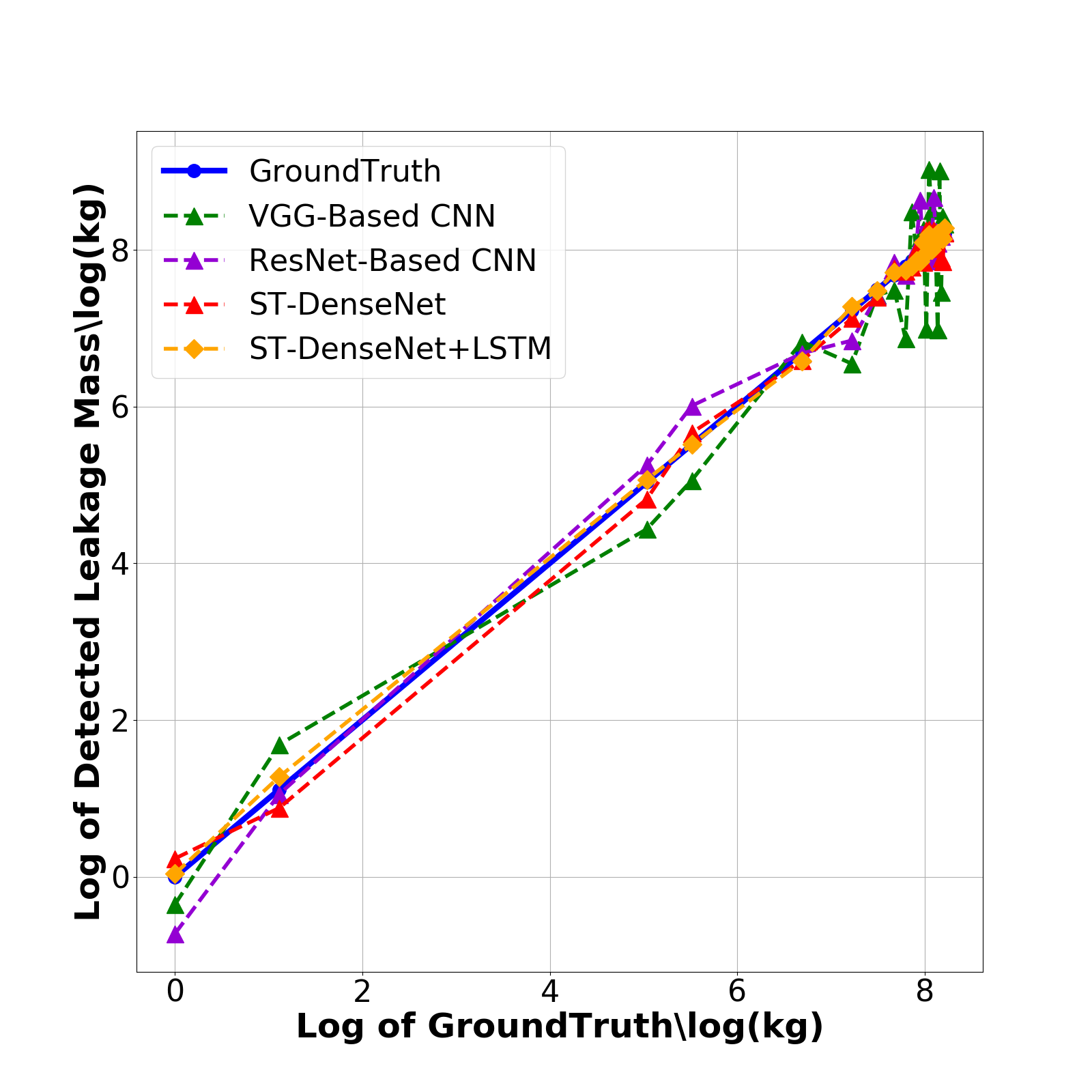}}
\subfigure[]{\includegraphics[width=0.5\linewidth]{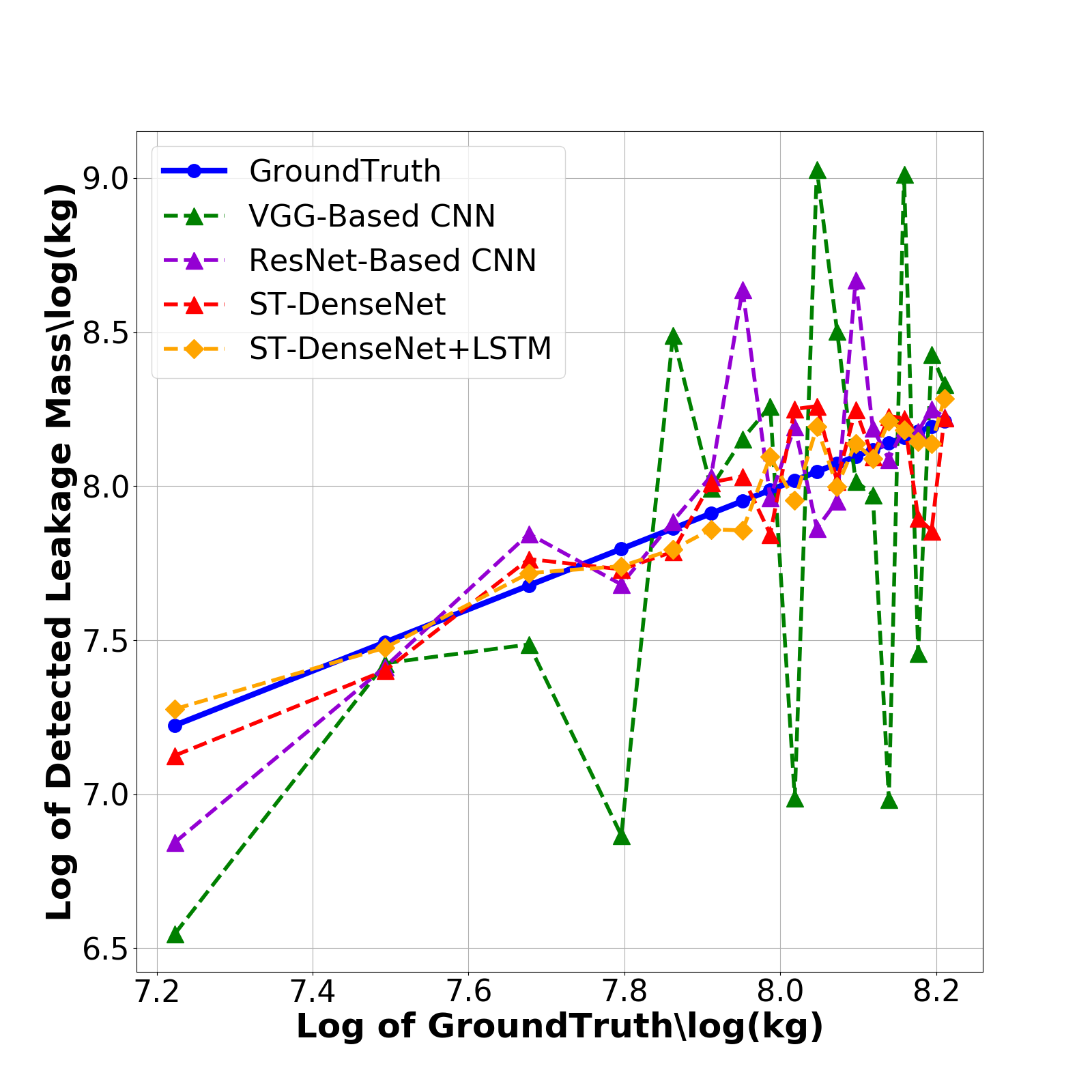}}}
\caption{Illustration of four methods to detect CO$_2$ leakage mass at location 1~km away from injector. Figure~\ref{fig:groundtruth_vs_detected_pre_1km}(a) shows the detection results in ``Log of GroundTruth VS. Log of Detected Mass'' of the 200-year leakage process, and Fig.~\ref{fig:groundtruth_vs_detected_pre_1km}(b) is the enlarged view of the last 19 detection results in 200 years. We provide detection results using four network architectures including VGGNet~(green), ResNet~(purple), ST-DenseNet~(red), and ST-DenseNet with LSTM~(orange). The ground-truth is in blue. We observe that ST-DenseNet+LSTM model yields the most accurate detection results and the minimum variance among all four methods.}
\label{fig:groundtruth_vs_detected_pre_1km}
\end{figure}

\begin{figure*}
\centerline{
\subfigure[]{\includegraphics[width=0.50\linewidth]{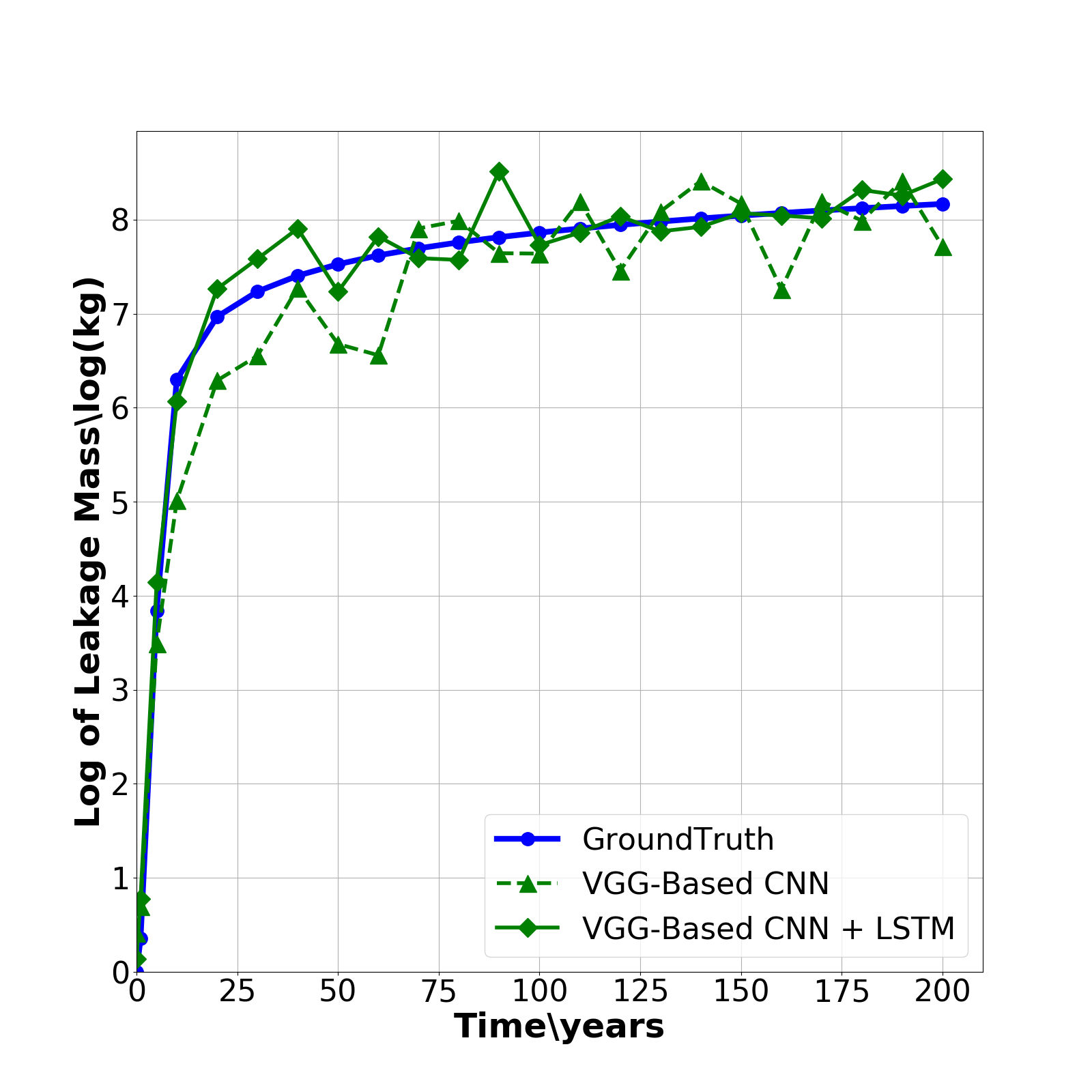}}
\subfigure[]{\includegraphics[width=0.50\linewidth]{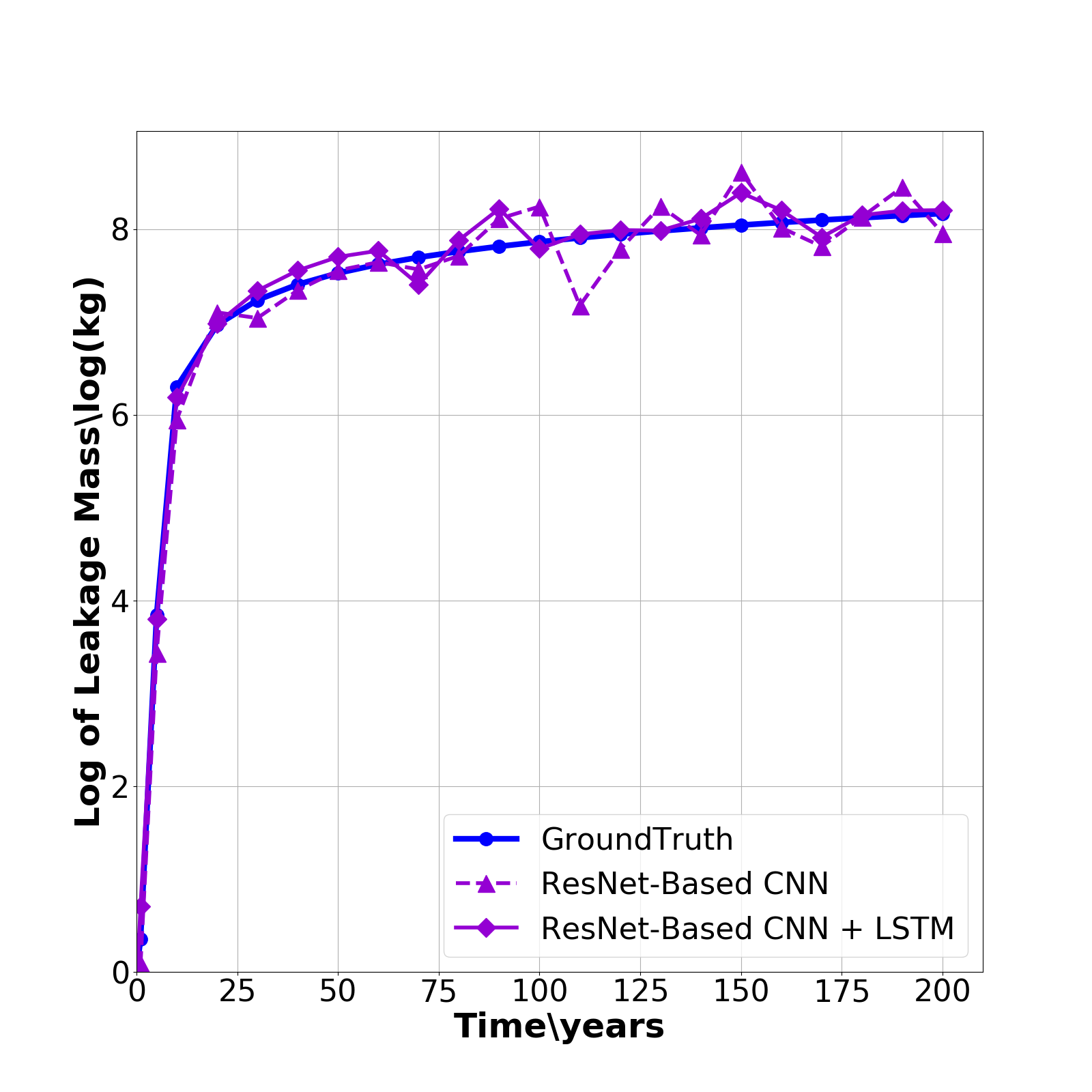}}}
\centerline{
\subfigure[]
{\includegraphics[width=0.50\linewidth]{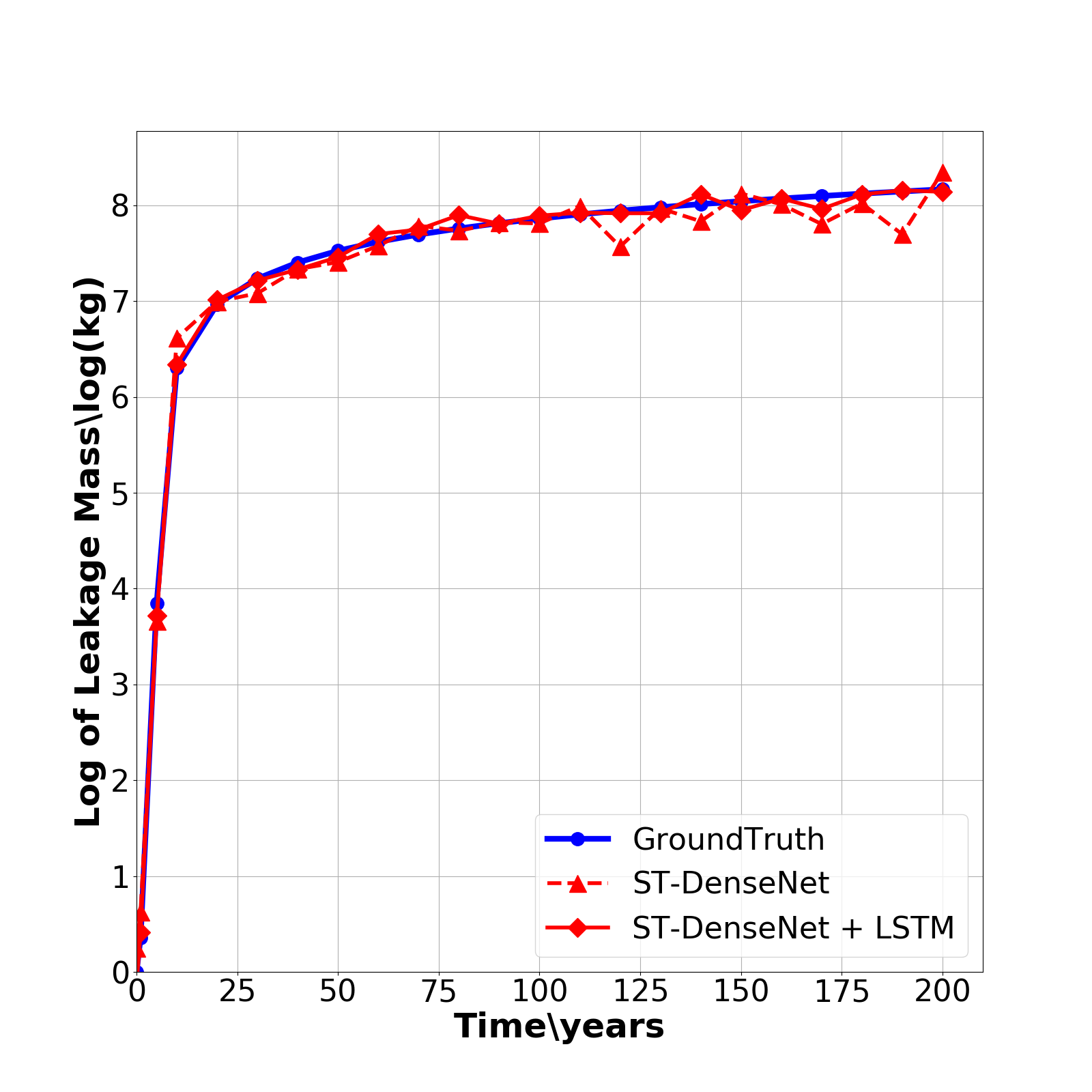}}
}
\caption{Illustration of three mthods to detect CO$_2$ leakage mass at location 3~km away from injector. We provide detection results using three network architectures including VGGNet~(a), ResNet~(b), and ST-DenseNet~(c). Within each sub-figure, we also provide detection results using network with LSTM (in ``$-\diamond-$'') and without LSTM (in ``$-\triangle-$''). The ground-truth is plotted in blue.We observe that (1) the incorporation of LSTM improves the detection accuracy for all three CNN architectures including ST-DenseNet; (2) ST-DenseNet yields the most accurate detection results of all three methods.}
\label{fig:exp_detection_mass2}
\end{figure*}

\begin{figure}
\centering
\centerline{
\subfigure[]{\includegraphics[width=0.5\linewidth]{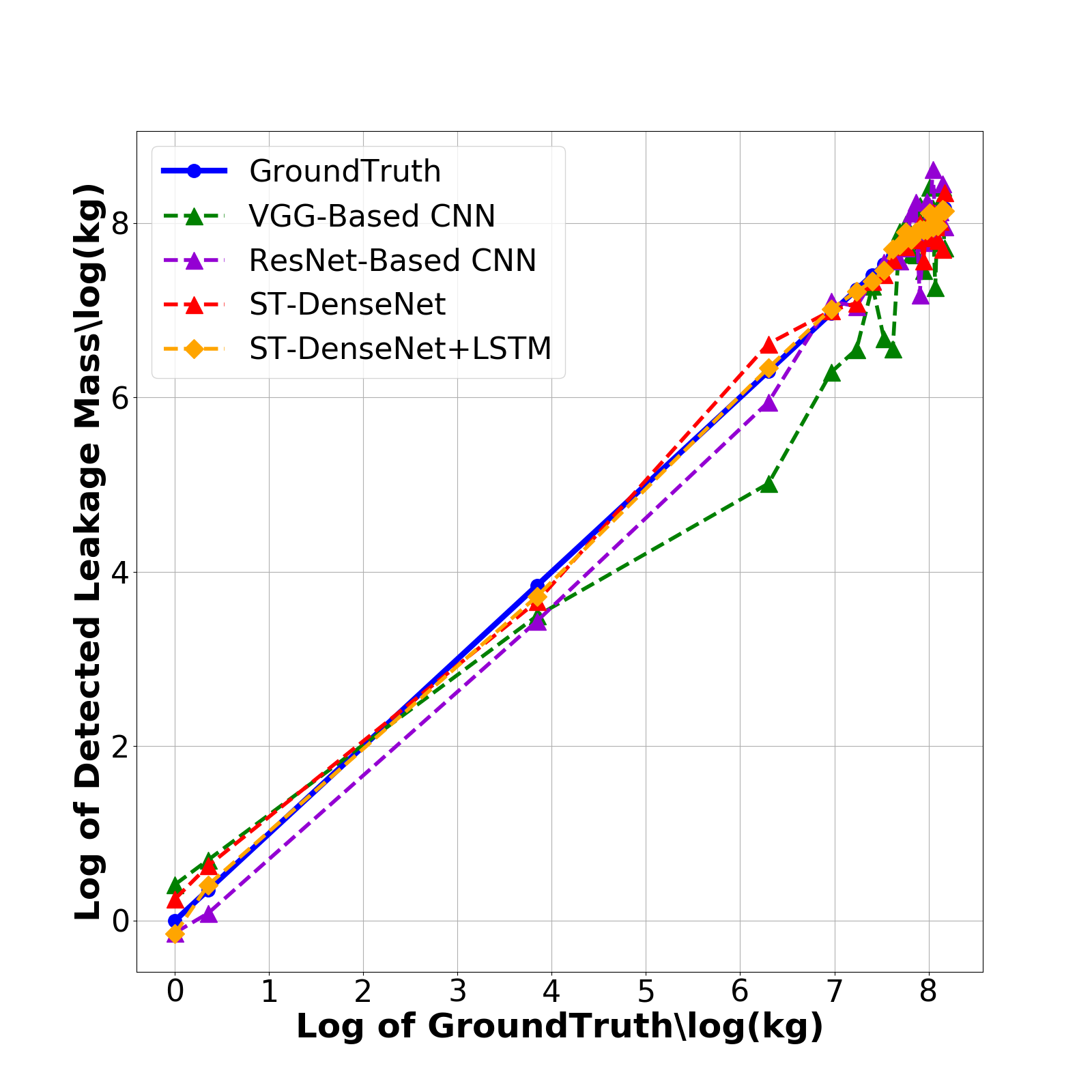}} 
\subfigure[]{\includegraphics[width=0.5\linewidth]{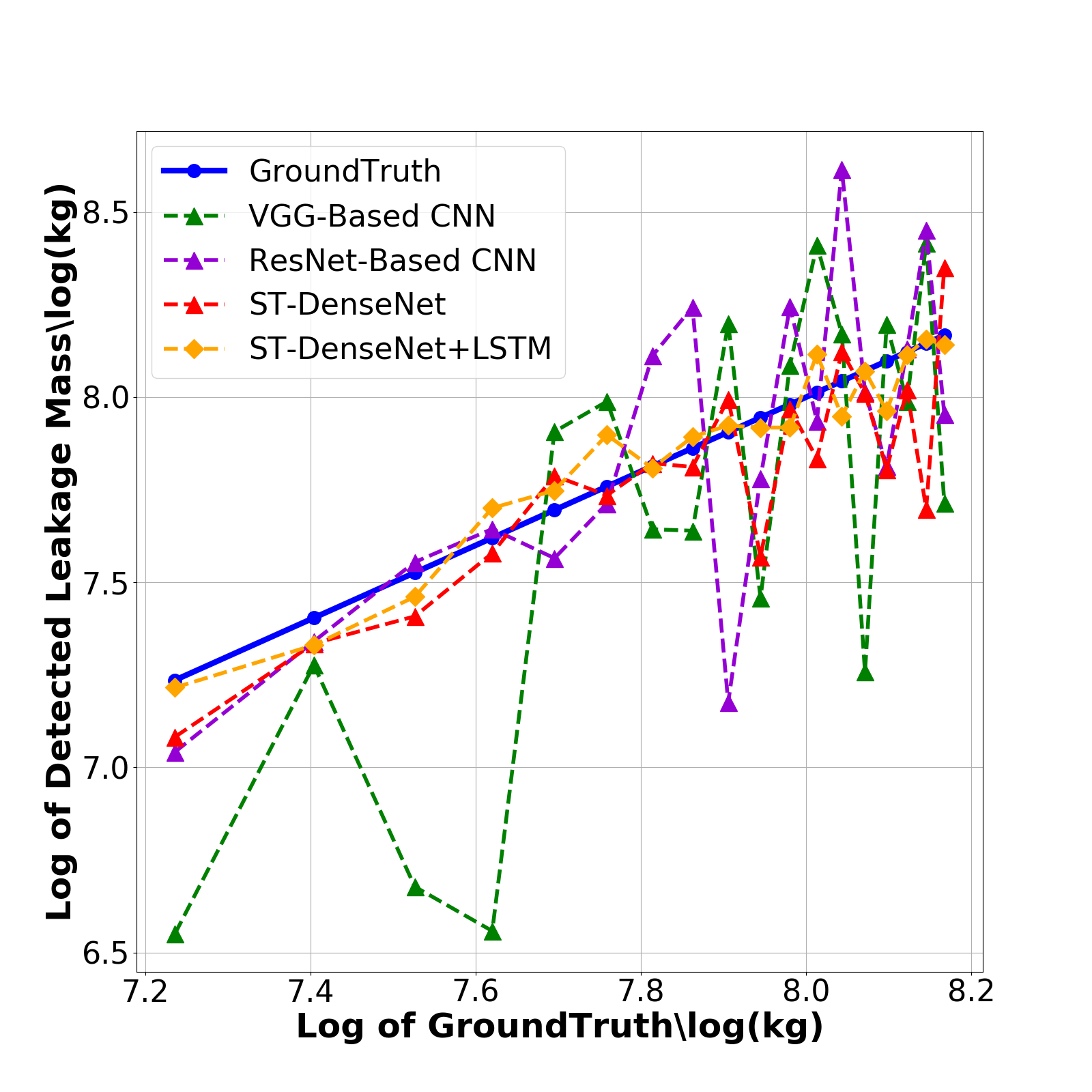}}}
\caption{Illustration of four methods to detect CO$_2$ leakage mass at location 3~km away from injector. Figure~\ref{fig:groundtruth_vs_detected_pre_3km}(a) shows the detection results in ``Log of GroundTruth VS. Log of Detected Mass'' of the 200-year leakage process, and Fig.~\ref{fig:groundtruth_vs_detected_pre_3km}(b) is the enlarged view of the last 19 detection results in 200 years. We provide detection results using four network architectures including VGGNet~(green), ResNet~(purple), ST-DenseNet~(red), and ST-DenseNet with LSTM~(orange). The ground-truth is in blue. We observe that ST-DenseNet+LSTM model yields the most accurate detection results and the minimum variance among all four methods.}
\label{fig:groundtruth_vs_detected_pre_3km}
\end{figure}

\begin{figure*}
\centerline{
\subfigure[]{\includegraphics[width=0.50\linewidth]{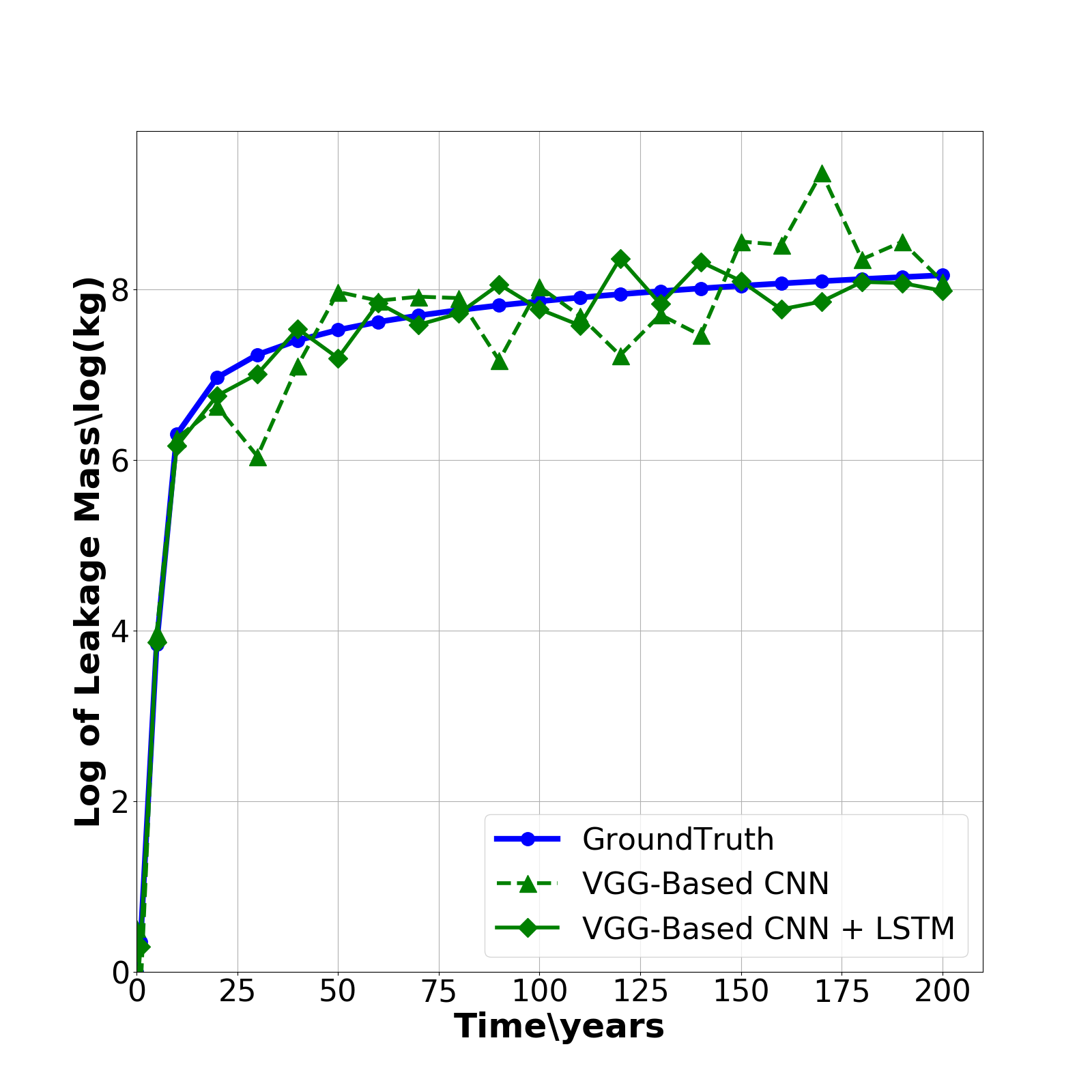}}
\subfigure[]{\includegraphics[width=0.50\linewidth]{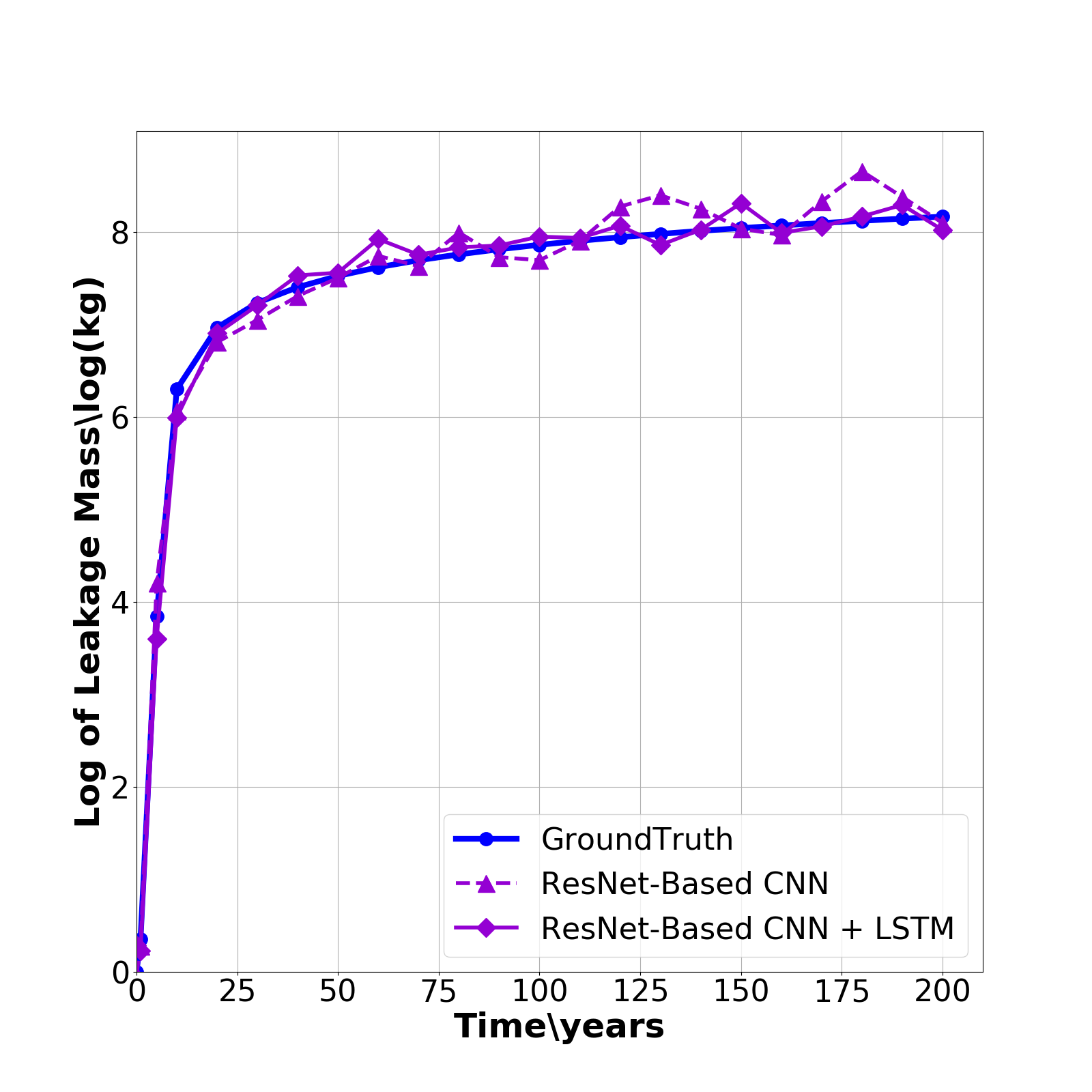}}}
\centerline{
\subfigure[]
{\includegraphics[width=0.50\linewidth]{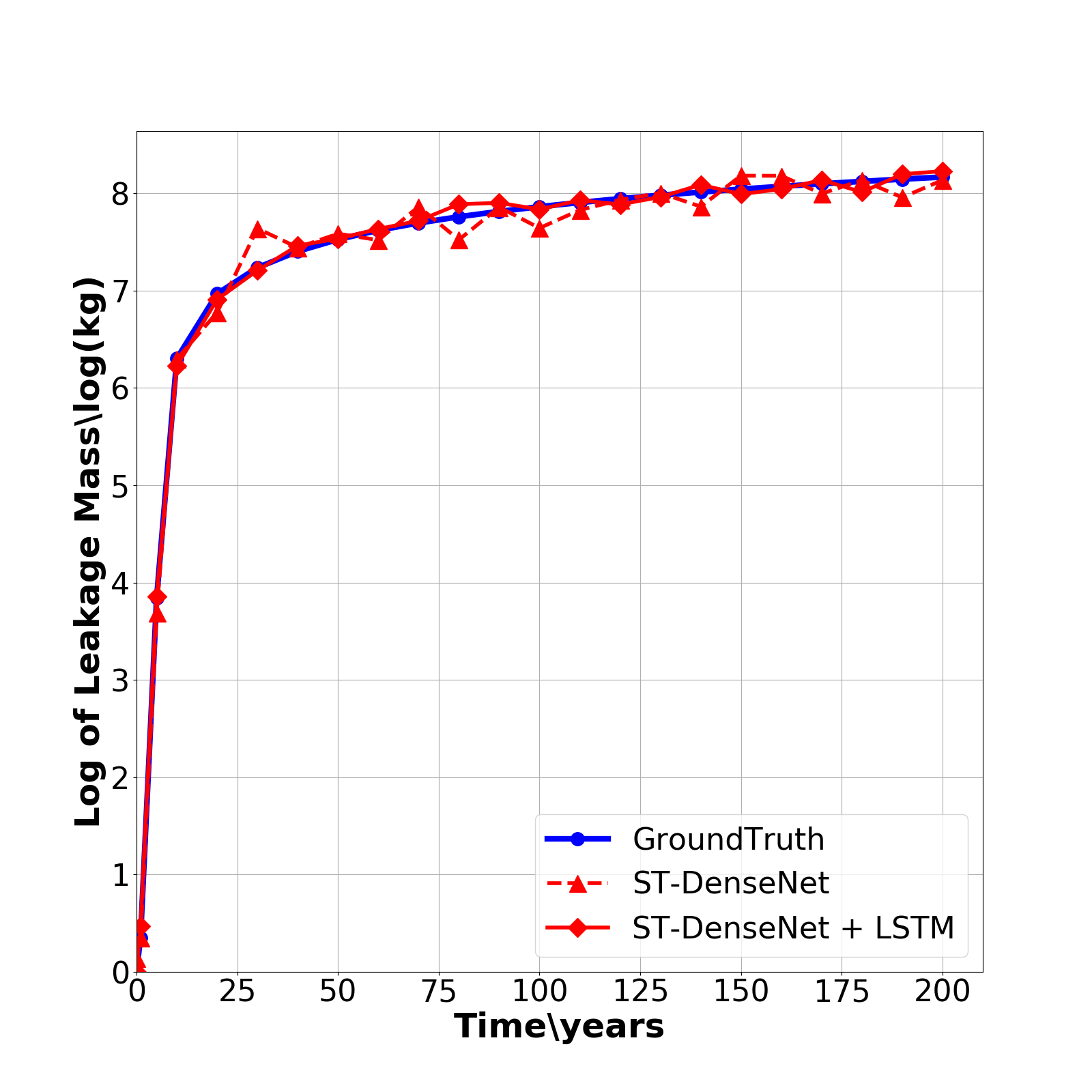}}
}
\caption{ Illustration of three mthods to detect CO$_2$ leakage mass at location 6~km away from injector. We provide detection results using three network architectures including VGGNet~(a), ResNet~(b), and ST-DenseNet~(c). Within each sub-figure, we also provide detection results using network with LSTM (in ``$-\diamond-$'') and without LSTM (in ``$-\triangle-$''). The ground-truth is plotted in blue.We observe that (1) the incorporation of LSTM improves the detection accuracy for all three CNN architectures including ST-DenseNet; (2) ST-DenseNet yields the most accurate detection results of all three methods.}
\label{fig:exp_detection_mass3}
\end{figure*}

\begin{figure}
\centering
\centerline{
\subfigure[]{\includegraphics[width=0.5\linewidth]{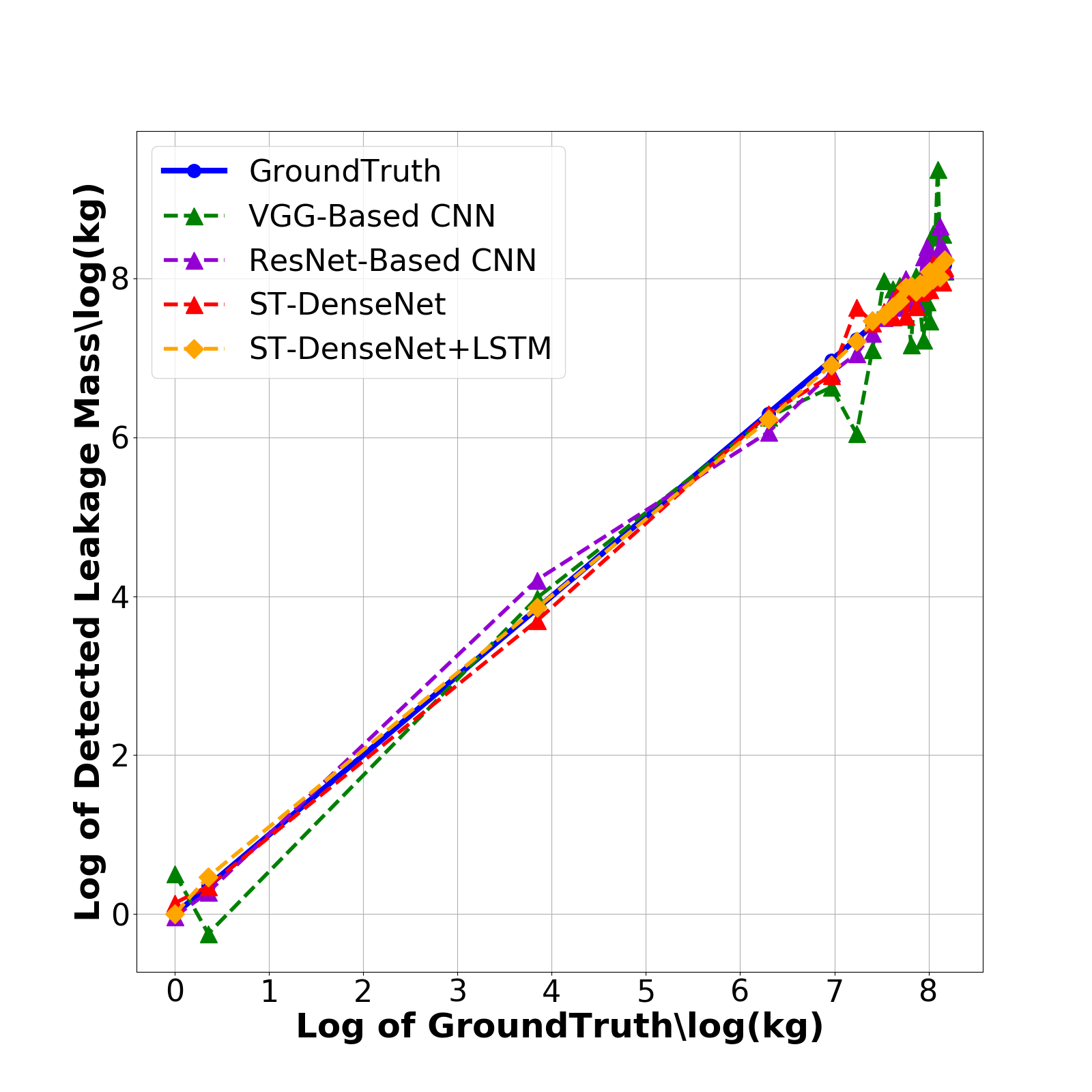}} 
\subfigure[]{\includegraphics[width=0.5\linewidth]{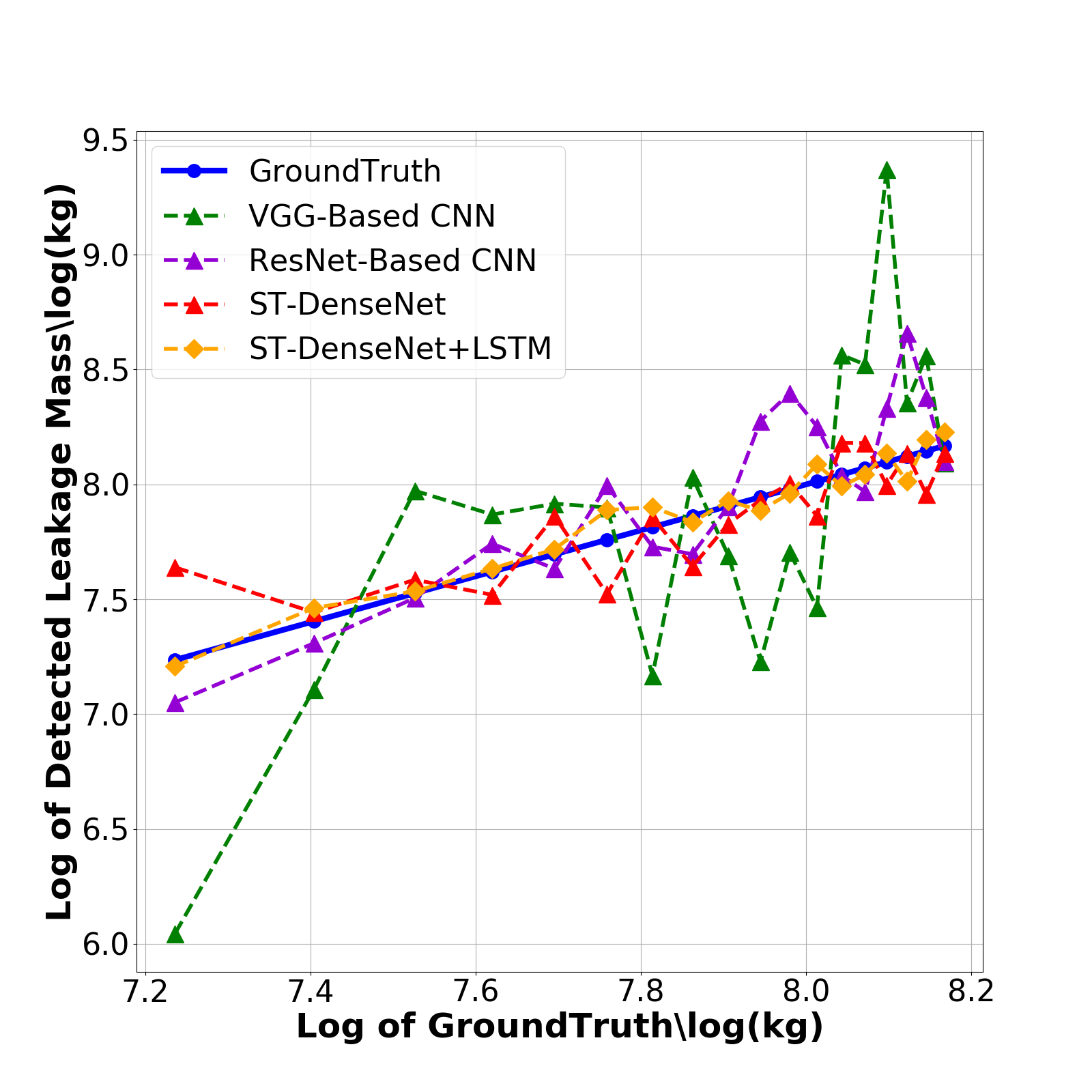}}}
\caption{Illustration of four methods to detect CO$_2$ leakage mass at location 6~km away from injector. Figure~\ref{fig:groundtruth_vs_detected_pre_6km}(a) shows the detection results in ``Log of GroundTruth VS. Log of Detected Mass'' of the 200-year leakage process, and Fig.~\ref{fig:groundtruth_vs_detected_pre_6km}(b) is the enlarged view of the last 19 detection results in 200 years. We provide detection results using four network architectures including VGGNet~(green), ResNet~(purple), ST-DenseNet~(red), and ST-DenseNet with LSTM~(orange). The ground-truth is in blue. We observe that ST-DenseNet+LSTM model yields the most accurate detection results and the minimum variance among all four methods.}
\label{fig:groundtruth_vs_detected_pre_6km}
\end{figure}

\subsection{Test on Sequential Leakage Monitoring}
\label{Time-Sequential}

We also tested the monitoring situation when a sequence of seismic datasets are acquired over time. We selected three leakage scenarios from three locations shown in Fig.~\ref{fig:CO2-injector} and report their detection results from Figs.~\ref{fig:exp_detection_mass1} to \ref{fig:groundtruth_vs_detected_pre_6km}. To account for the dependency from sequential time series, we incorporated LSTM into ST-DenseNet as illustrated in Fig.~\ref{fig:ST-DensNet}. We chose VGGNet and ResNet as baseline methods. In order to compare fairly, we also incorporated LSTM with both VGGNet and ResNet, which result in 5 different CNN-based baseline models including VGG-Based CNN, VGG-Based CNN with LSTM, ResNet-Based CNN, ResNet-Based CNN with LSTM, and ST-DenseNet without LSTM. To illustrate the time dependency, we plot the detection results versus time obtained using VGG-based CNN~(Fig.~\ref{fig:exp_detection_mass1}(a), Fig.~\ref{fig:exp_detection_mass2}(a), and Fig.~\ref{fig:exp_detection_mass3}(a)), ResNet-based CNN~(Fig.~\ref{fig:exp_detection_mass1}(b), Fig.~\ref{fig:exp_detection_mass2}(b), and Fig.~\ref{fig:exp_detection_mass3}(b)), and ST-DenseNet~(Fig.~\ref{fig:exp_detection_mass1}(c), Fig.~\ref{fig:exp_detection_mass2}(c), and Fig.~\ref{fig:exp_detection_mass3}(c)). We also show the results of the detection results of the leakage mass at 22 different times using all 6 methods including ST-DenseNet. In each of these figures, the ground-truth is plotted in blue. The results using VGG-Based CNNs are plotted in dark green~(Fig.~\ref{fig:exp_detection_mass1}(a), Fig.~\ref{fig:exp_detection_mass2}(a), and Fig.~\ref{fig:exp_detection_mass3}(a)). The results using ResNet-Based CNNs are plotted in purple~(Fig.~\ref{fig:exp_detection_mass1}(b), Fig.~\ref{fig:exp_detection_mass2}(b), and Fig.~\ref{fig:exp_detection_mass3}(b)). The results using ST-DenseNet are plotted in red~(Fig.~\ref{fig:exp_detection_mass1}(c), Fig.~\ref{fig:exp_detection_mass2}(c), and Fig.~\ref{fig:exp_detection_mass3}(c)). 

In Fig.~\ref{fig:exp_detection_mass1}, we observe that the incorporation of LSTM improves the detection accuracy for all three CNN architectures including ST-DenseNet. In particular, the CNN networks without LSTM not only yielded detections with significant oscillations over time, but also produced inaccurate leakage mass. On the other hand, by incorporating LSTM, the detection results using all three CNN networks were significantly improved. The  oscillation issue have been alleviated and the detection accuracy was also improved. This demonstrates that 
accounting the dependency from sequential time series is critical in generating an accurate leakage detection model. Furthermore, by comparing all three versions of CNN networks with LSTM, we notice that ST-DenseNet still yields the most accurate results as can be observed by comparing Fig.~\ref{fig:exp_detection_mass1}(a), ~\ref{fig:exp_detection_mass1}(b), and ~\ref{fig:exp_detection_mass1}(c). Similar performance can be observed from Figs.~\ref{fig:exp_detection_mass2} and \ref{fig:exp_detection_mass3}. 

To illustrate the detection accuracy, we also provide the plots of Log of GroundTruth VS Log of Detected Mass using VGG-Based CNN (green), ResNet-Based CNN (purple), ST-DenseNet (red), and ST-DenseNet + LSTM (orange) for three different leakage locations at 1~km (Fig.~\ref{fig:groundtruth_vs_detected_pre_1km}), 3~km(Fig.~\ref{fig:groundtruth_vs_detected_pre_3km}) and 6~km(Fig.~\ref{fig:groundtruth_vs_detected_pre_6km}), respectively. We observe that the detection results of ST-DenseNet with LSTM  are more accurate than those obtained using all the other detection methods with the smallest variances from GroundTruth. Therefore, we conclude from this test that ST-DenseNet with LSTM yields more accurate detection results than those obtained using VGGNet with/without LSTM and ResNet with/without LSTM when a time sequence of data is available.

\subsection{Tests on Robustness}

Robustness is an important issue for any deep neural network~\citep{Fawzi-2017-Robustness}. In this section, we investigate the robustness of ST-DenseNet by implementing (1) an intra-site cross-location test to demonstrate the generalization ability of our model on unknown test data, and (2) a noisy data test to validate the performance of our method under noisy environment.
   
\subsubsection{Intra-Site Cross-Location Test}

We tested ST-DenseNet on intra-site cross-location datasets to demonstrate its robustness and weak generalization. The samples in our dataset were collected from three different locations within a site. In particular, the distances of the three locations to CO$_2$ injector were 1~km, 3~km, and 6~km, respectively~(shown in Figure~\ref{fig:CO2-injector}). We provide the key parameters used for generating simulations at the three well locations in Table~\ref{table:ParameterForDifferentWells}. To further visualize the leakage parameters at those three locations in different time, we provide the pressure change in Table~\ref{table:ParameterForDifferentWells_Pressure}~(Appendix~A) and $\mathrm{CO}_2$ saturation in Table~\ref{table:ParameterForDifferentWells_Saturation}~(Appendix~B).

\begin{table}
\centering
\begin{tabular}{ |c|c|c|c| }
\hline
Parameters & \multicolumn{3}{|c|}{Well Location at 1km, 3km, 6km} \\ 
\hline
\hline
\multirow{ 5}{*}{Layer Depths} & \multicolumn{3}{|c|}{Layer~1: Atmosphere (1e-30~m thick)} \\
 & \multicolumn{3}{|c|}{Layer 2: Upper Caprock (10~m thick))}  \\
 & \multicolumn{3}{|c|}{Layer 3: Etchegoin Formation (536.23~m thick)}  \\
& \multicolumn{3}{|c|}{Layer 4: Macoma-Chanac Formation (679.04~m thick)} \\
 & \multicolumn{3}{|c|}{Layer 5: Santa Margarita-McLure Formation (185.94~m thick)}  \\
\hline
Temperature & \multicolumn{3}{|c|}{$40^\circ C$}\\
\hline
Porosity         & \multicolumn{3}{|c|}{$0.35$}\\
\hline
\end{tabular}
\caption{Comparison of key parameters at the three well locations. }
\label{table:ParameterForDifferentWells}
\end{table}

We used samples from 2 locations for training and tested the detection methods on the third, resulting in three different combinations of training-testing groups. We then used the average prediction accuracies obtained using ST-DenseNet based on the three groups as the overall accuracy. Similar to previous tests, all the accuracies were calculated under the criterion of "$\pm 5\%$ Accuracy". This intra-site cross-location test can be much more challenging than all of the previous tests for two reasons. From the physics point of view, in all previous tests, the datasets used for training and testing came from physics simulations of the conditions. Therefore, ST-DenseNet captured the physical correspondence between data and target values (leakage mass in our problem). On the other hand, in this robustness test, training sets and testing sets were obtained from physics simulations based on different conditions. Therefore, the relationship that was learned from training may or may not have fully represented the one in the test sets. From the mathematical point of view, the goal for  any machine learning algorithm including deep learning is to find a model to fit data. With both training sets and testing sets drawn from the same distribution~(previous tests), it becomes relatively less challenging to generate a model comparing to the situations when training sets and testing sets are drawn from two different distributions~(current test). 

Table \ref{table:results2} presents the results of the robustness test for the various methods. The use of different locations for training and testing did degrade the accuracy compared with tests using the same locations for training and testing (Table \ref{table:results1}). This indicates that there is some discrepancy between the training sets and testing sets. However, ST-DenseNet still achieves reasonable detection results and yields the highest accuracy among all the CNN-based networks in this intra-site cross-location test. As illustrated in the last row of Table~\ref{table:results2}, by incorporating  LSTM into ST-DenseNet, we can improve the detection accuracy further.

\begin{table}
\centering
\begin{tabular}{ |c|c|c|c|c| }
\hline
Accuracy of Different Test Location& 1km & 3km & 6km & Overall Acc \\ 
\hline
\hline
VGG-based CNN& 68.6\% & 73.3\% & 72.1\% & 71.1\% \\
\hline
ResNet-based CNN& 73.6\% & 77.4\% & 79.9\% & 76.7\% \\
\hline
ST-DenseNet& 78.4\% & 81.9\% & 81.5\% & 80.0\% \\
\hline
ST-DenseNet + LSTM& \textbf{79.3\%} & \textbf{82.3\%} & \textbf{82.6\%} & \textbf{81.2\%} \\
\hline
\end{tabular}
\caption{The robustness test on generelization using different models including VGG-based CNN, ResNet-based CNN, ST-DenseNet and ST-DenseNet+LSTM model. All the accuracy are calculated under the criterion of "$\pm 5\%$ Accuracy". Our ST-DenseNet outperforms all the other CNN-based models. By incorporating LSTM structure and using the sequential information, the detection results are further improved.}
\label{table:results2}
\end{table}

\begin{table}
\centering
\begin{tabular}{ |c|c|c|c|c| c|}
\hline
MSE of Different Dection Methods& Without Noise & With Noise  & Degradation~(With Noise - Without Noise) \\ 
\hline
\hline
VGG-based CNN& 6.891 & 9.485  & 2.594\\
\hline
ResNet-based CNN& 3.972 & 5.784 & 1.812\\
\hline
ST-DenseNet& 1.136 & 2.653  & 1.494\\
\hline
ST-DenseNet + LSTM& \textbf{0.765} & \textbf{1.327}  & \textbf{0.562} \\
\hline
\end{tabular}
\caption{A summary of the MSE values of VGG-based CNN, ResNet-based CNN, ST-DenseNet and ST-DenseNet with LSTM architecture for testing data with or without noise. Our ST-DenseNet not only yields the smaller MSE error values, but also degrades the least among all three CNN networks. }
\label{table:MSE}
\end{table}

\begin{figure*}
\centerline{
\includegraphics[width=0.90\linewidth]{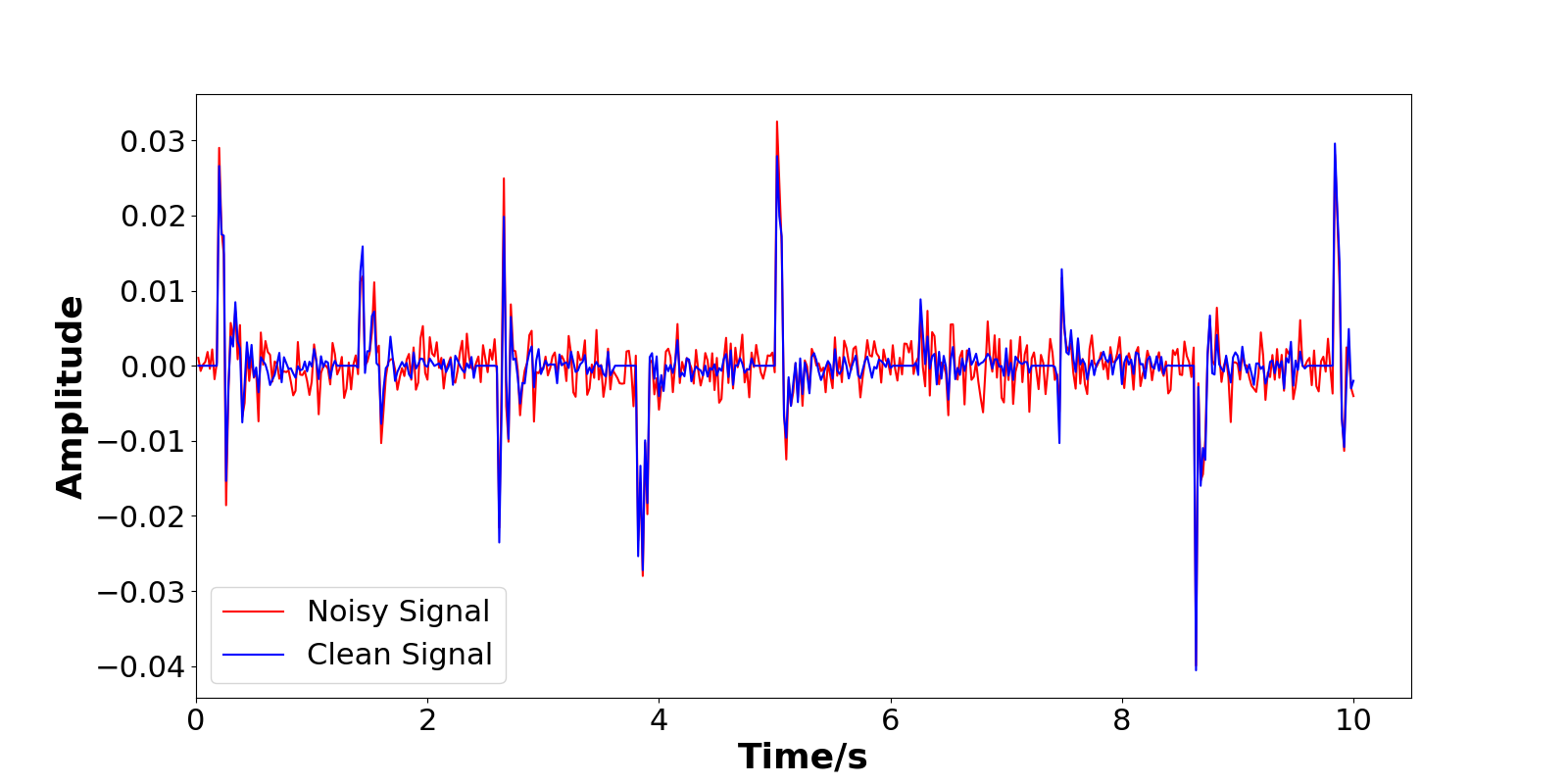}}
\caption{Illustration of the 1-D clean seismic data (blue) and the noisy signal with 30 db Gaussian noise (red).}
\label{fig:noise_signal}
\end{figure*}

\subsubsection{Noisy Data Test} 

The impact of random noise to deep networks is another important fact to consider for the robustness~\citep{Fawzi-2017-Robustness}. In real seismic measurements, random noise can be usually categorized as ``natural noise'' and ``cultural noise''~\citep{Seismic-2017-Li}. The level of the noise may be varied significantly due to differences in recording instruments, weather conditions, environment issues etc. In this test, we generated 30~db additive Gaussian noise, which can be a good approximation of the noise level that is seen near surface according to~\cite{comparison-2007-Young}. The Gaussian noise was added to the clean simulated seismic data as shown in Fig~\ref{fig:noise_signal}. To validate the performance of ST-DenseNet algorithms in different scenarios, we report in Table~\ref{table:MSE} to show the results on both random leakage monitoring and sequential leakage monitoring tests. We use mean squared error(MSE) as the metric to evaluate the performance of our detection methods under noisy environment. The MSE is defined as
\begin{equation}
\mathrm{MSE} = \frac{1}{n}\sum^{n}_{i=1}(Y_i-\hat{Y_i})^2,
\end{equation}
where $n$ is the total number of testing samples, $Y_i$ is the real leakage mass value, $\hat{Y_i}$ is the detected leakage mass value.

All CNN-based networks are degraded when noise is present in the data (comparing Column~3 to Column 2 in Table~\ref{table:MSE}). However, ST-DenseNet yields the smaller MSE error values with noise~(MSE $\approx$ 2.653 and MSE $\approx$ 1.327 with LSTM) comparing to VGGNet (MSE $\approx$ 9.485) and ResNet (MSE $\approx$ 5.784). Moreover, the last column in Table~\ref{table:MSE} shows that ST-DenseNet has the least degradation among all three CNN networks. 

\begin{figure}
\centering
\centerline{
\subfigure[]{\includegraphics[width=0.50\linewidth]{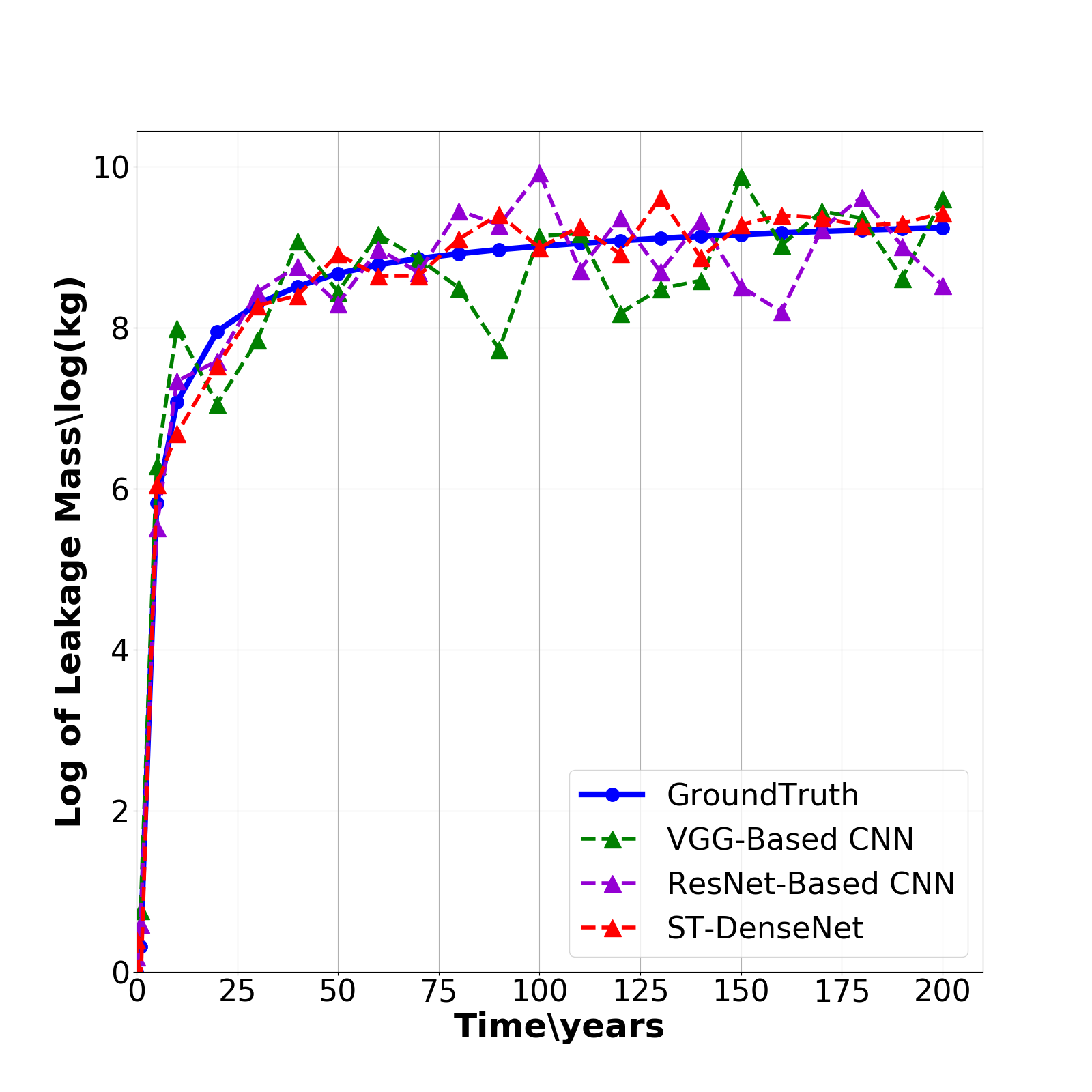}}
\subfigure[]{\includegraphics[width=0.50\linewidth]{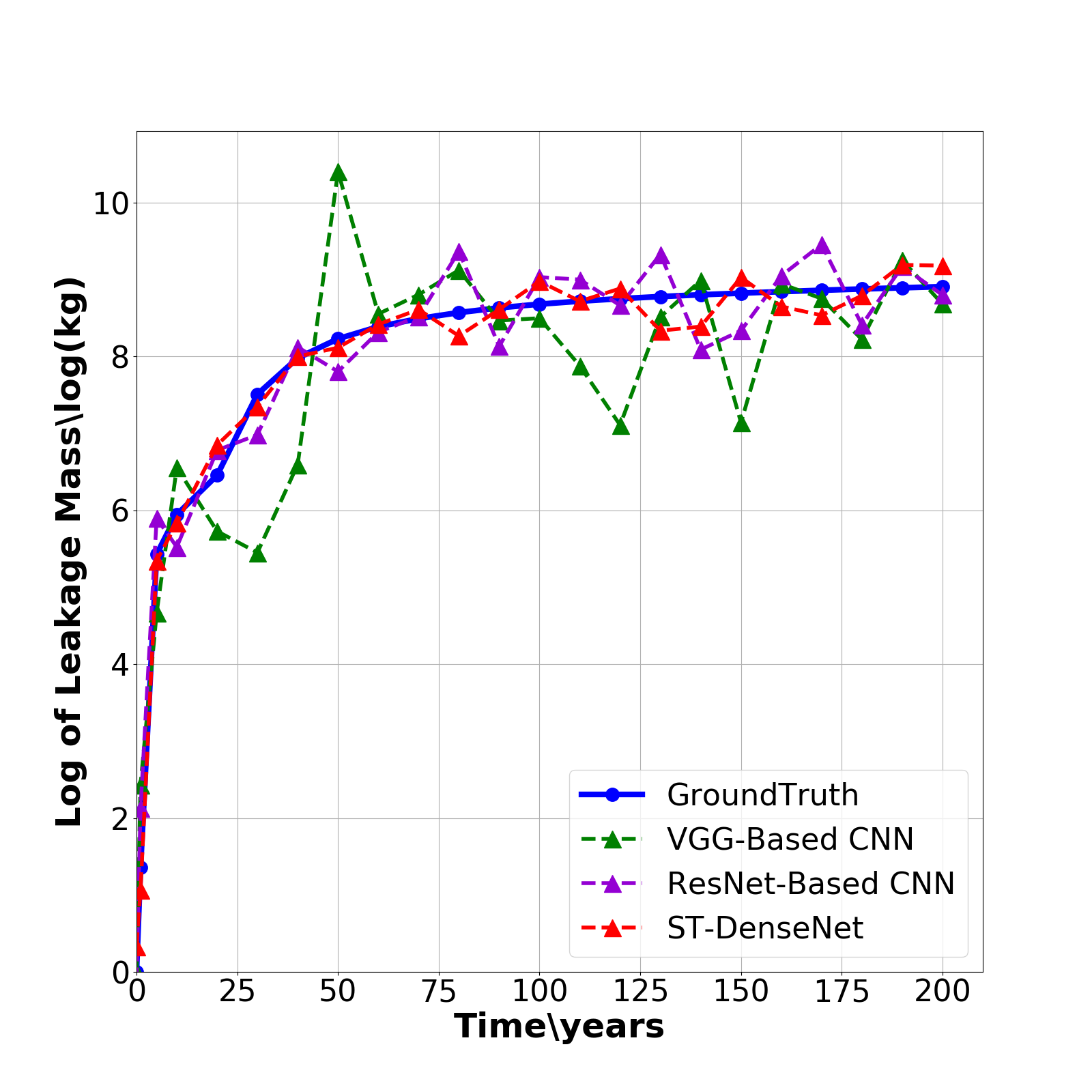}}}
\centerline{
\subfigure[]{\includegraphics[width=0.50\linewidth]{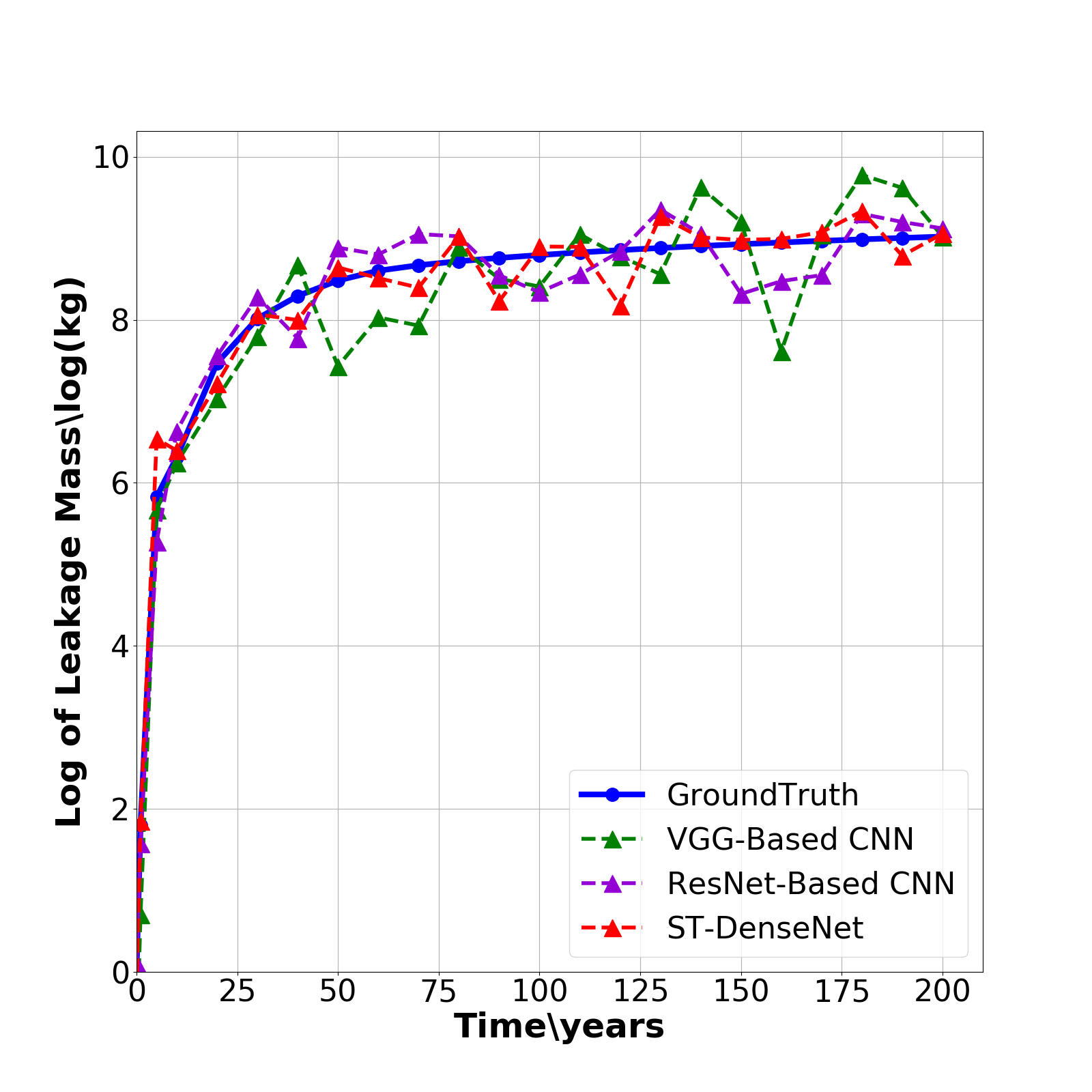}} 
\subfigure[]{\includegraphics[width=0.50\linewidth]{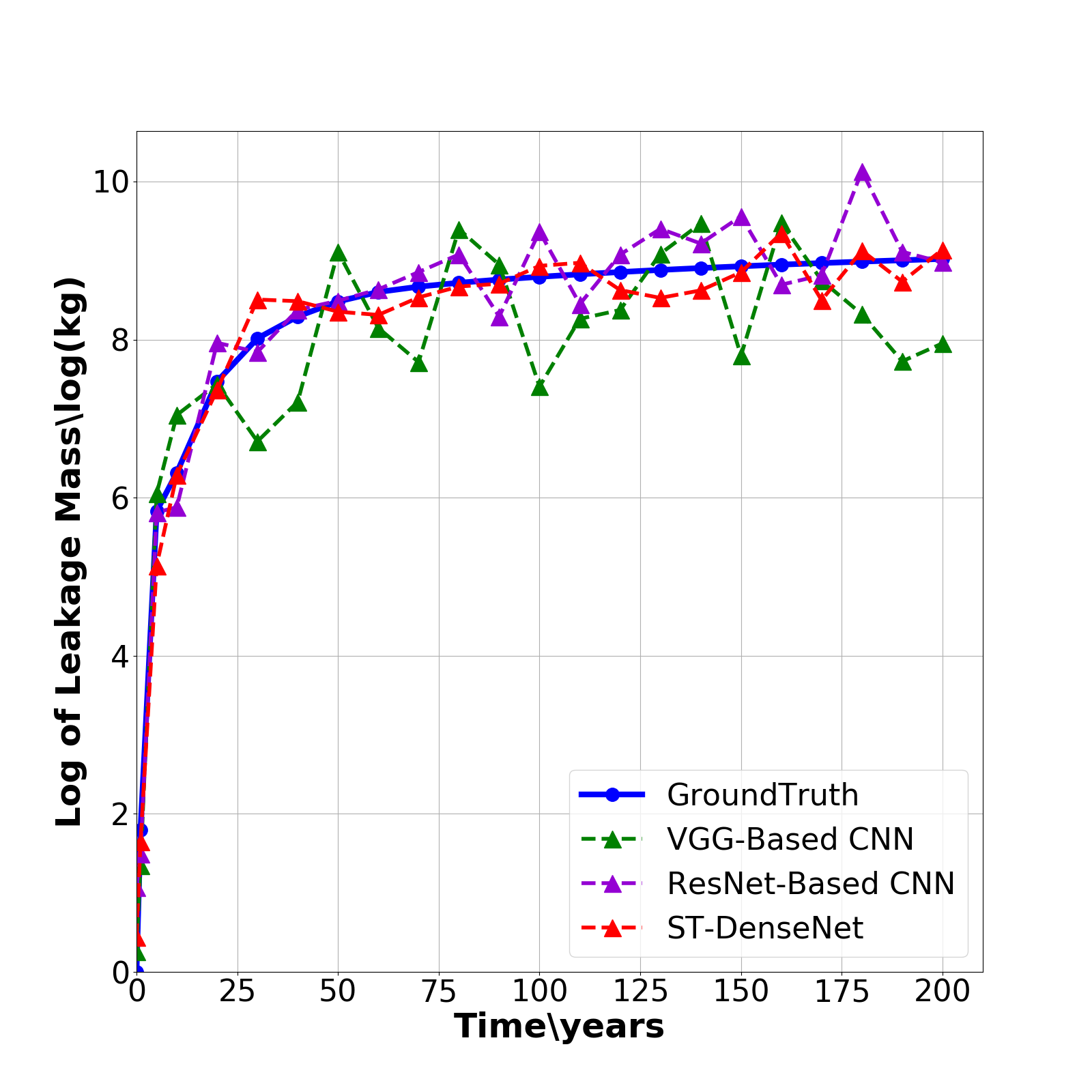}}}
\caption{Illustration of four CO$_2$ leakage mass detection scenarios on the data with 30db noise. Each of these sub-figures contains 22 different leakage mass over time (except the point of $t=0$).  The ground-truth is plotted in blue. We show results obtained using VGG-based CNN~(in green), ResNet-based CNN~(in purple), and ST-DenseNet~(in red). Our ST-DenseNet achieves most accurate CO$_2$ leakage mass detection results among the three CNN-based methods.}
\label{fig:noise_detection1}
\end{figure}

\begin{figure}
\centering
\centerline{
\subfigure[]{\includegraphics[width=0.50\linewidth]{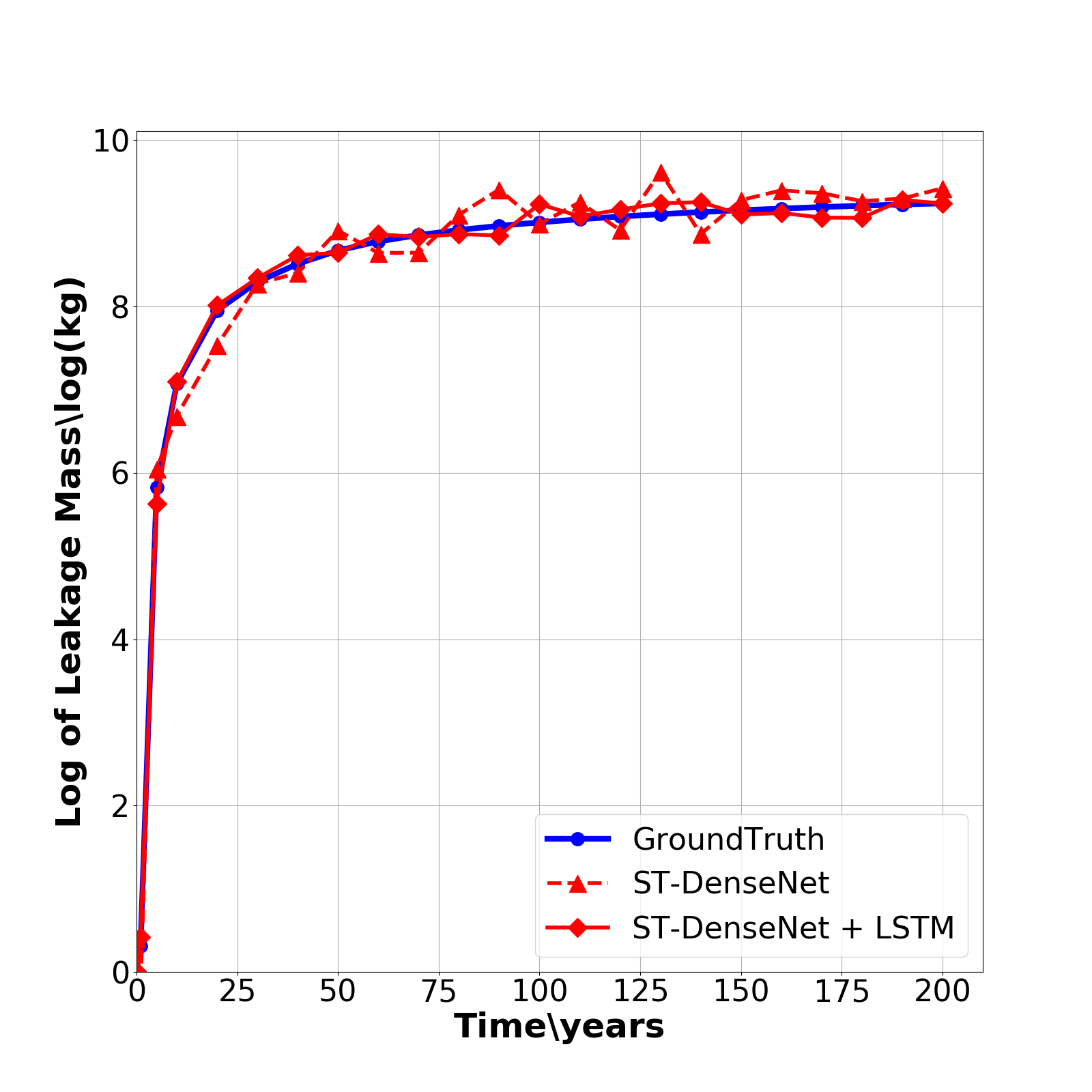}}
\subfigure[]{\includegraphics[width=0.50\linewidth]{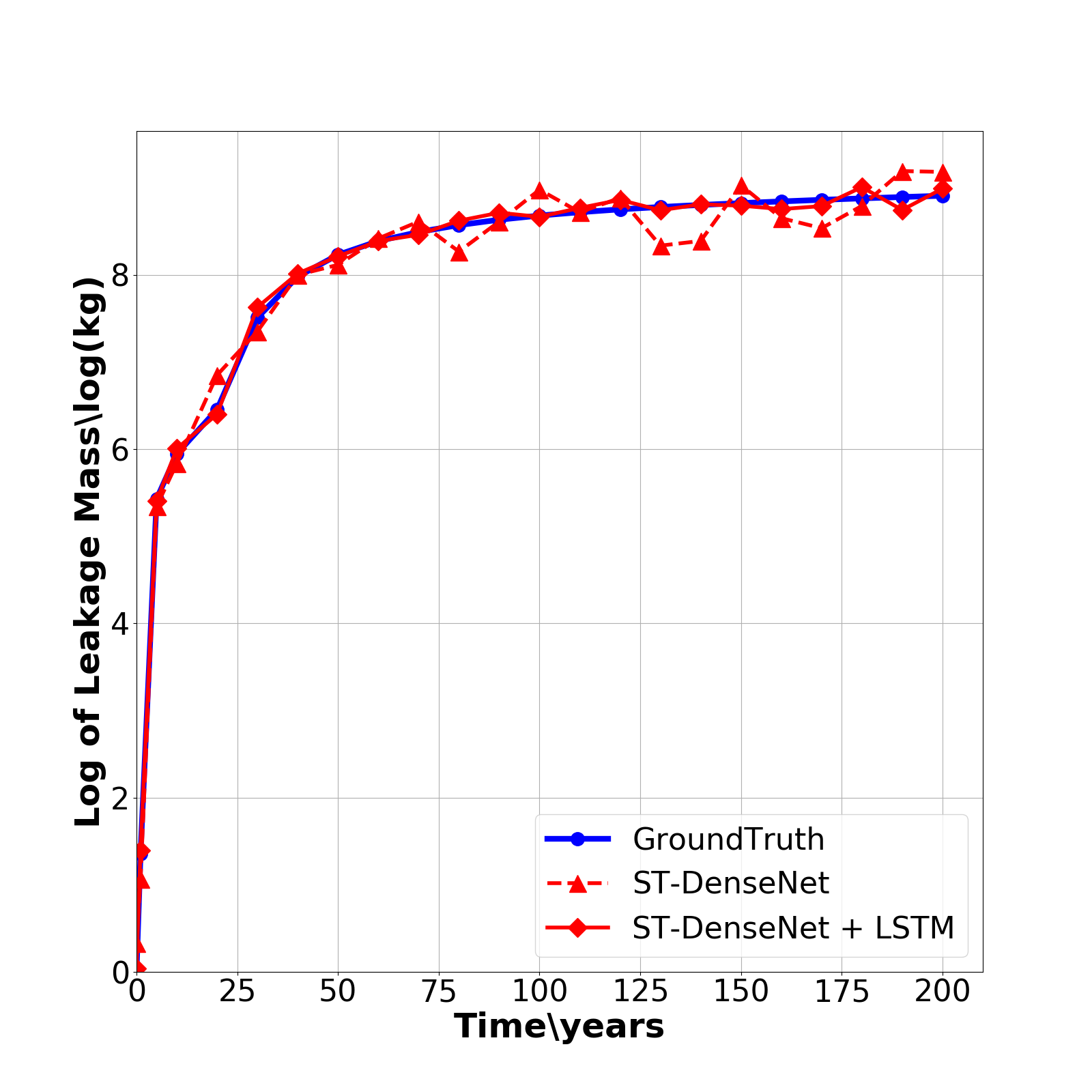}}}
\centerline{
\subfigure[]{\includegraphics[width=0.50\linewidth]{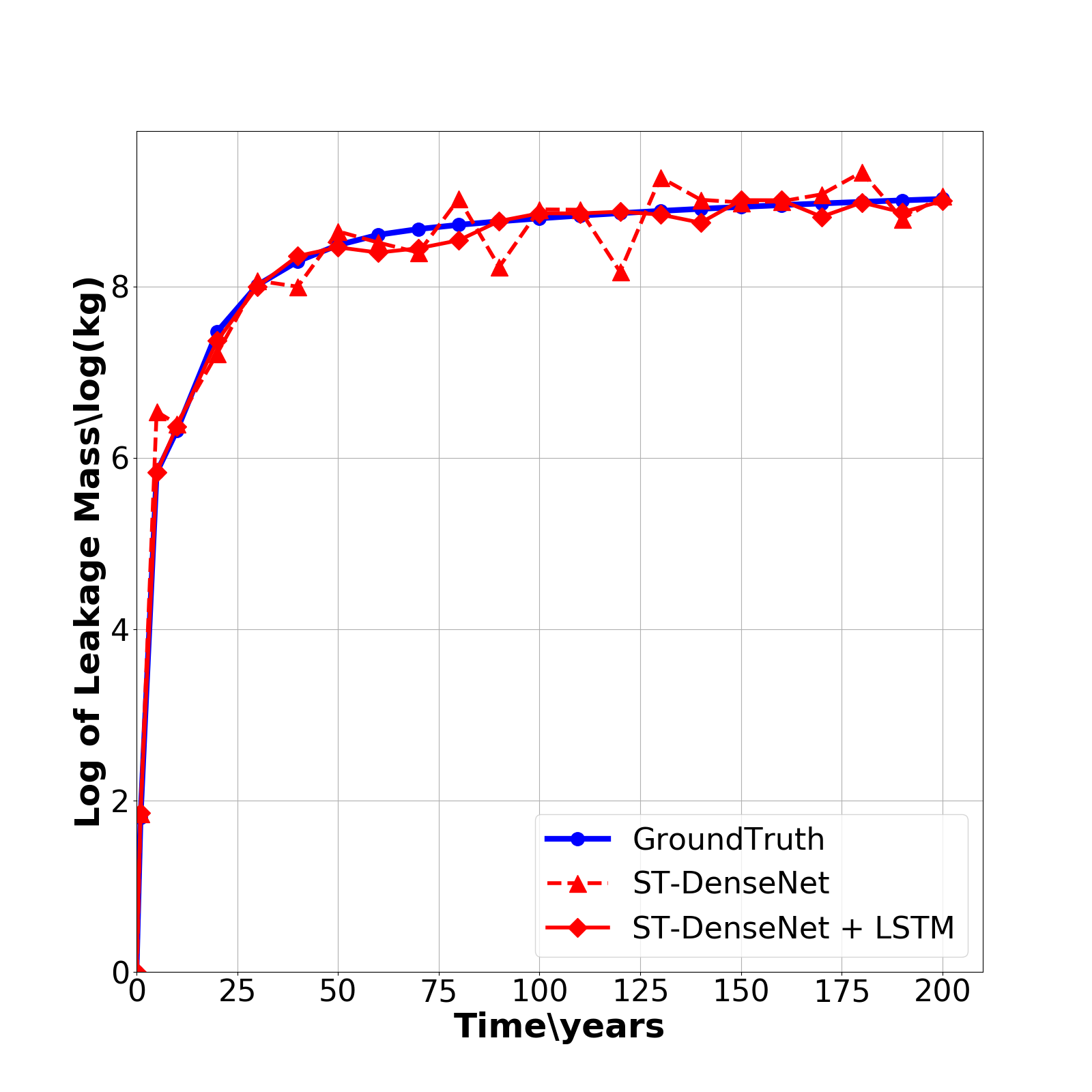}} 
\subfigure[]{\includegraphics[width=0.50\linewidth]{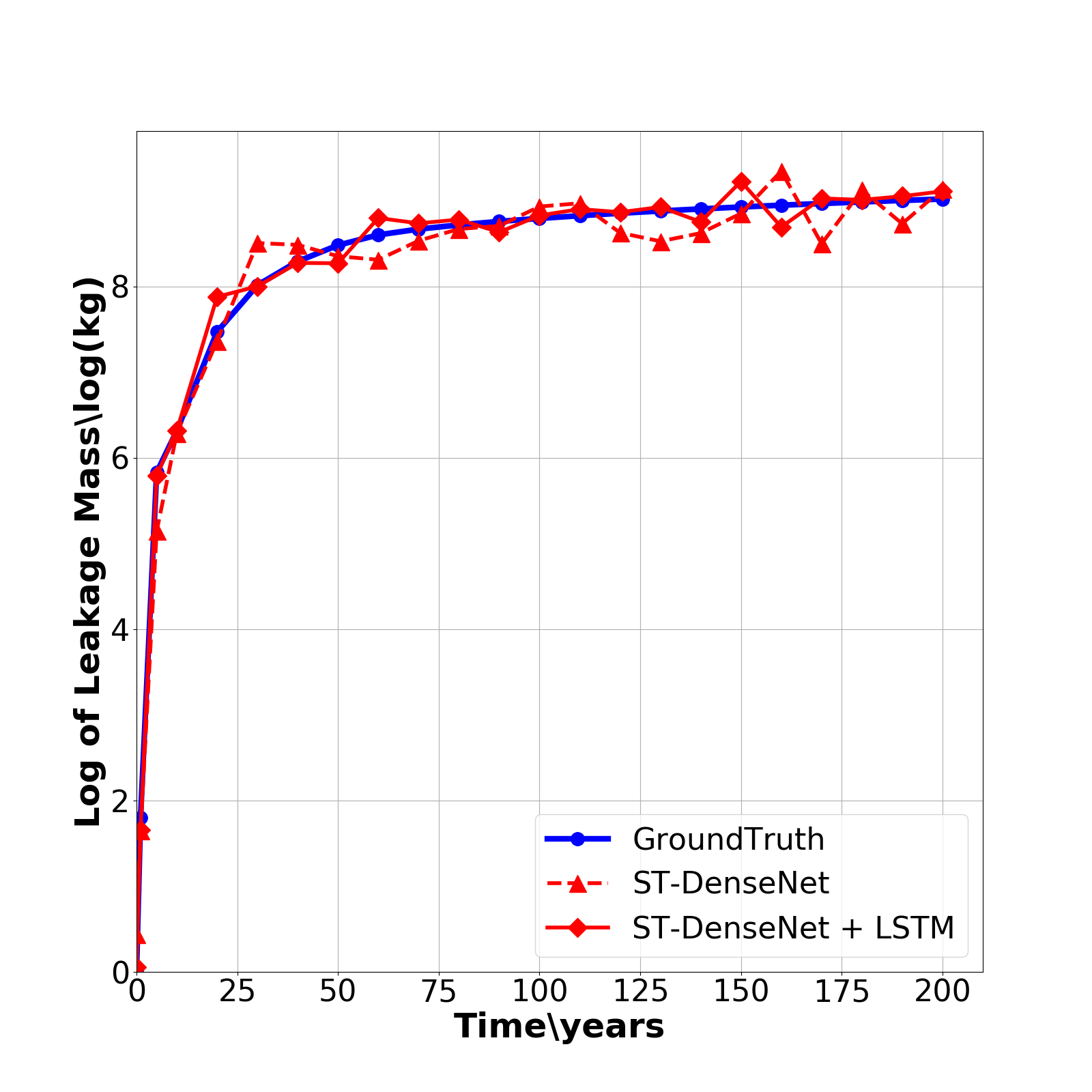}}}
\caption{Illustration of four CO$_2$ leakage mass detection scenarios on the data with 30db noise. Each of these sub-figures contains 22 different leakage mass over time (except the point of $t=0$).  The groundtruth is plotted in blue. We show results obtained using ST-DenseNet~(in ``$-\triangle-$'') and ST-DenseNet with LSTM~(in ``$-\diamond-$''). With the help from LSTM, the detection accuracy using ST-DenseNet can be much more improved. This again demonstrates that the dependency within the time sequence can be critical to improve the detection accuracy.}
\label{fig:noise_detection2}
\end{figure}

\begin{figure}
\centering
\centerline{
\subfigure[]{\includegraphics[width=0.5\linewidth]{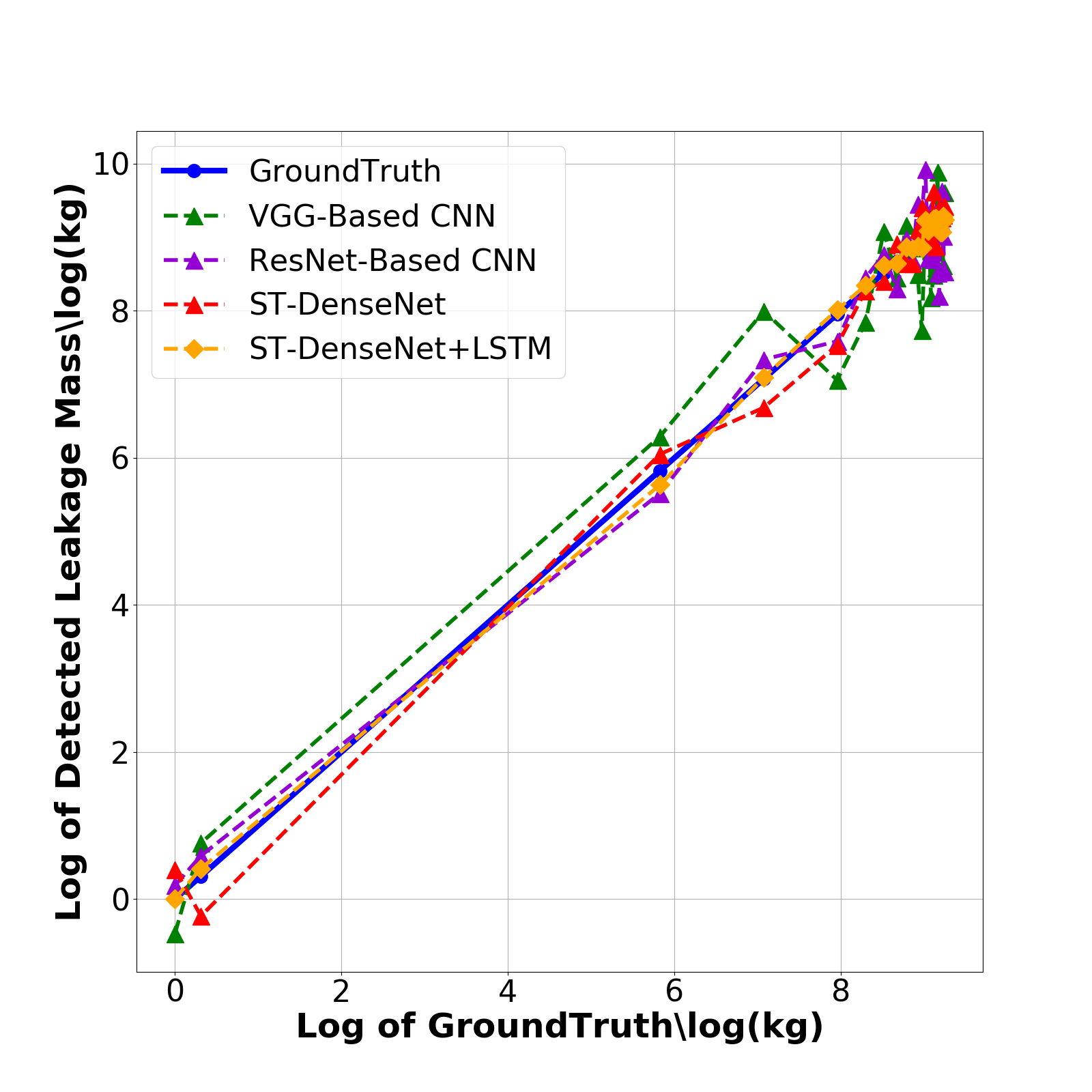}}
\subfigure[]{\includegraphics[width=0.5\linewidth]{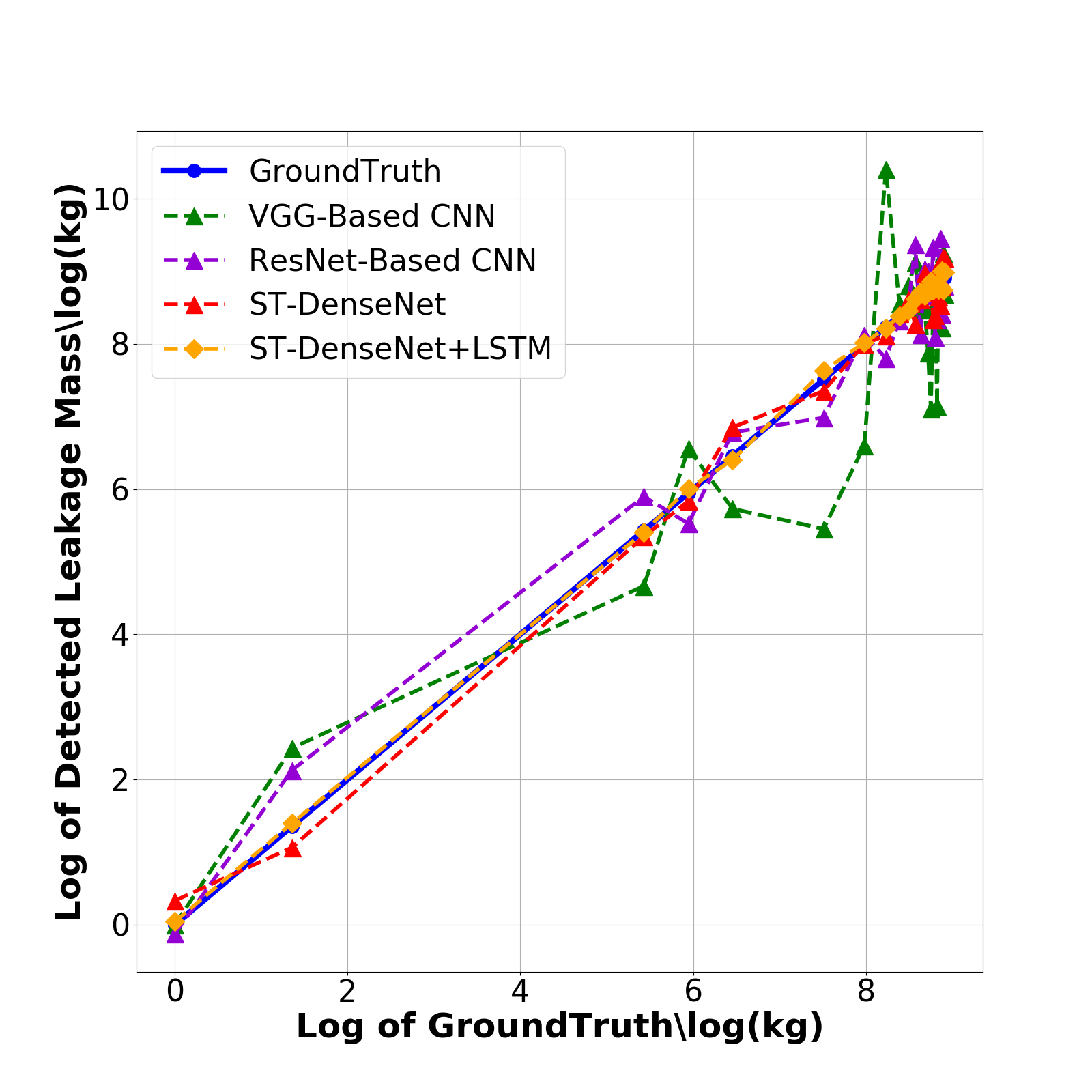}}}
\centerline{
\subfigure[]{\includegraphics[width=0.5\linewidth]{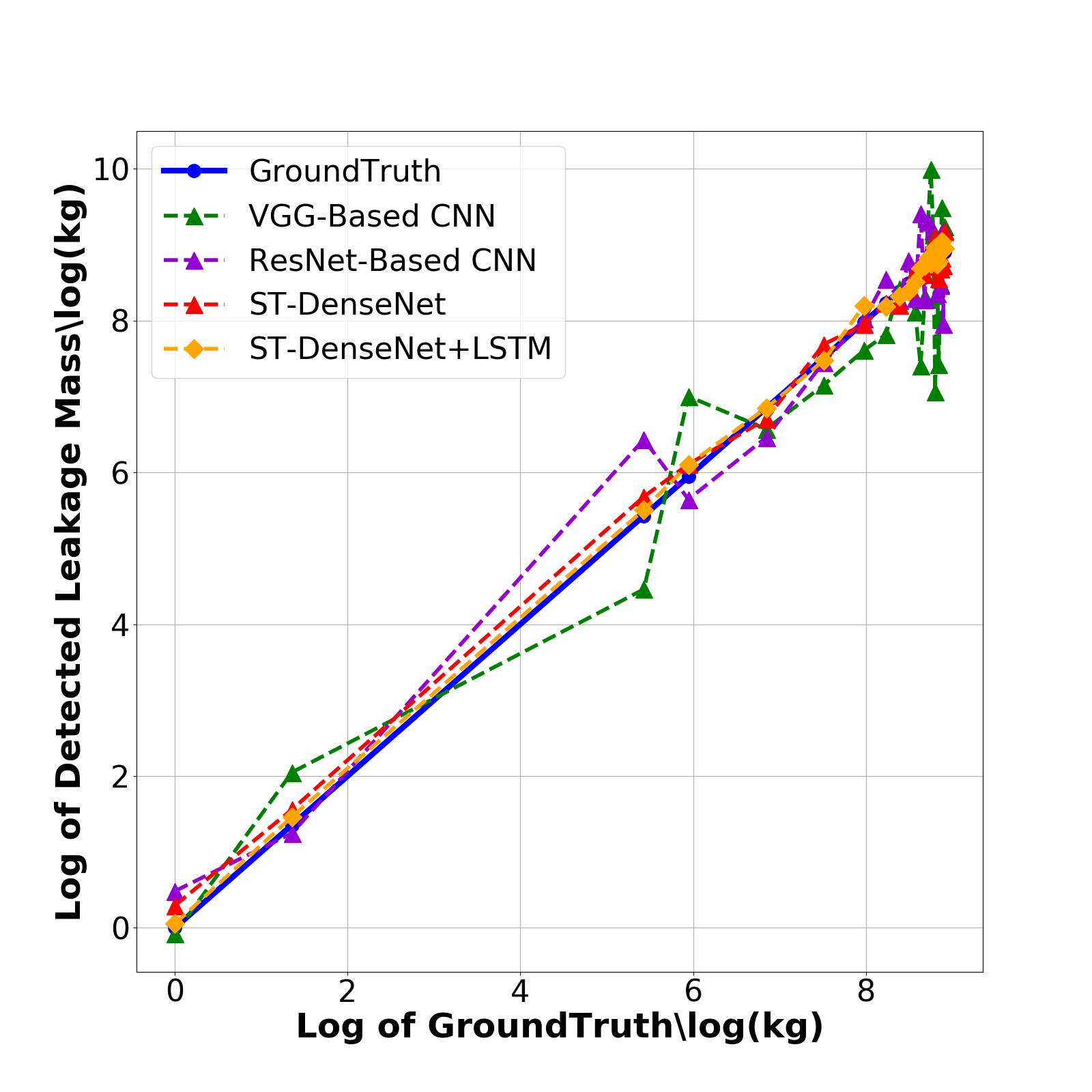}}
\subfigure[]{\includegraphics[width=0.5\linewidth]{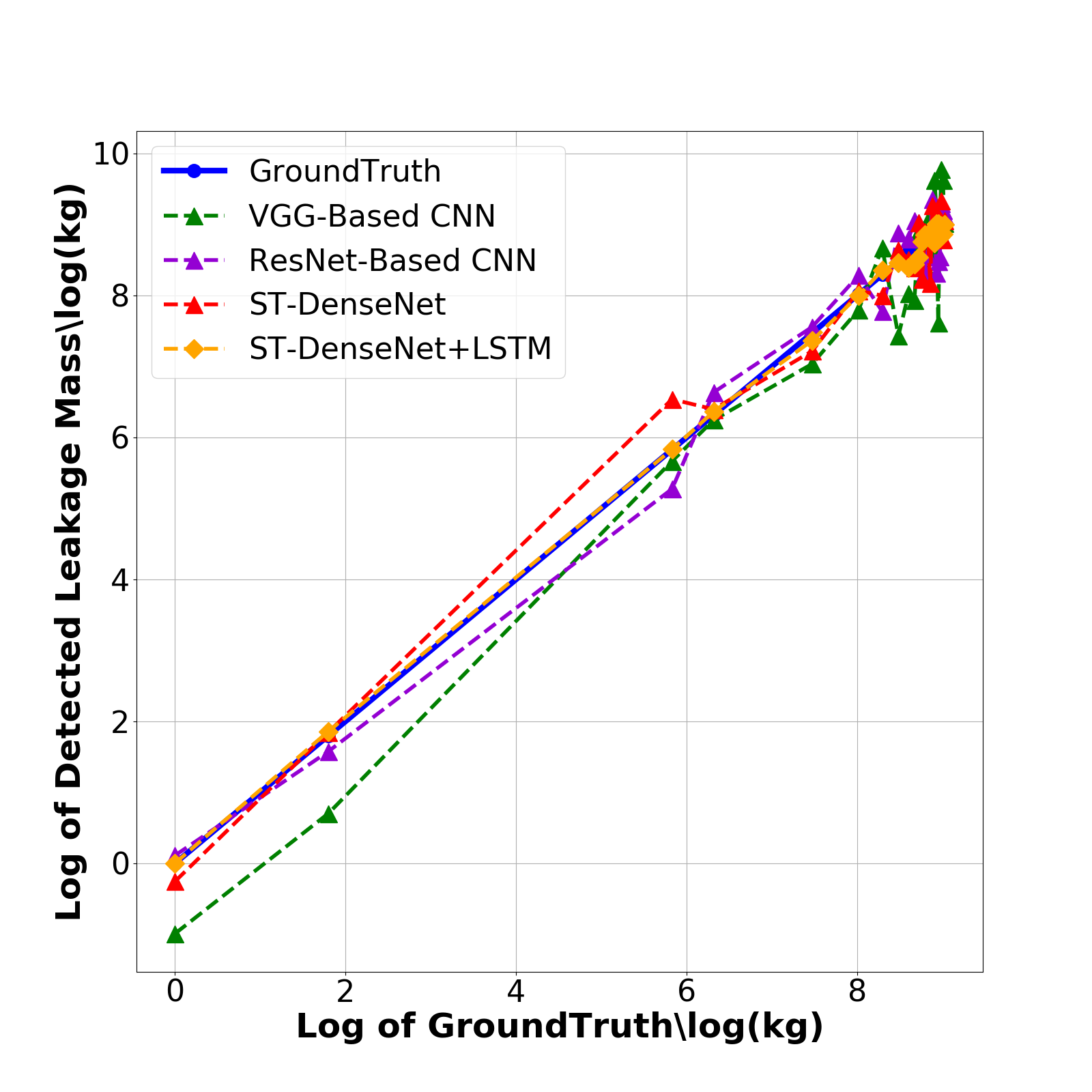}}}
\caption{Illustration of the detection results plotted in the Log of GroundTruth VS. Log of Detected Mass of the 200-year leakage process on the data with 30db noise. We provide detection results using four network architectures including VGGNet~(green), ResNet~(purple), ST-DenseNet~(red), and ST-DenseNet with LSTM~(orange). The ground-truth is in blue. We observe that ST-DenseNet+LSTM model yields the most accurate detection results and the minimum variance among all four methods.}
\label{fig:groundtruth_vs_detected_noise}
\end{figure}

\begin{figure}
\centering
\centerline{
\subfigure[]{\includegraphics[width=0.5\linewidth]{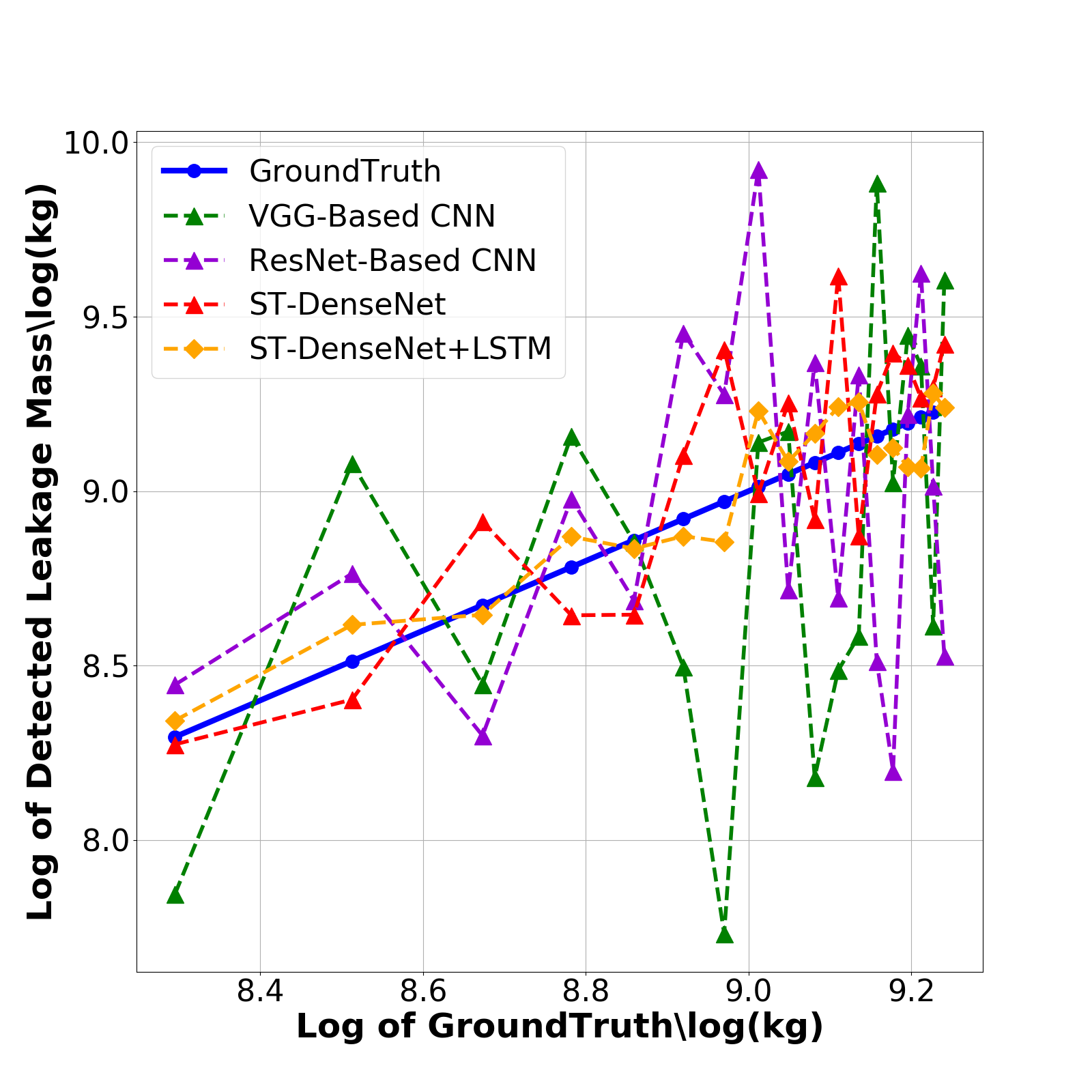}}
\subfigure[]{\includegraphics[width=0.5\linewidth]{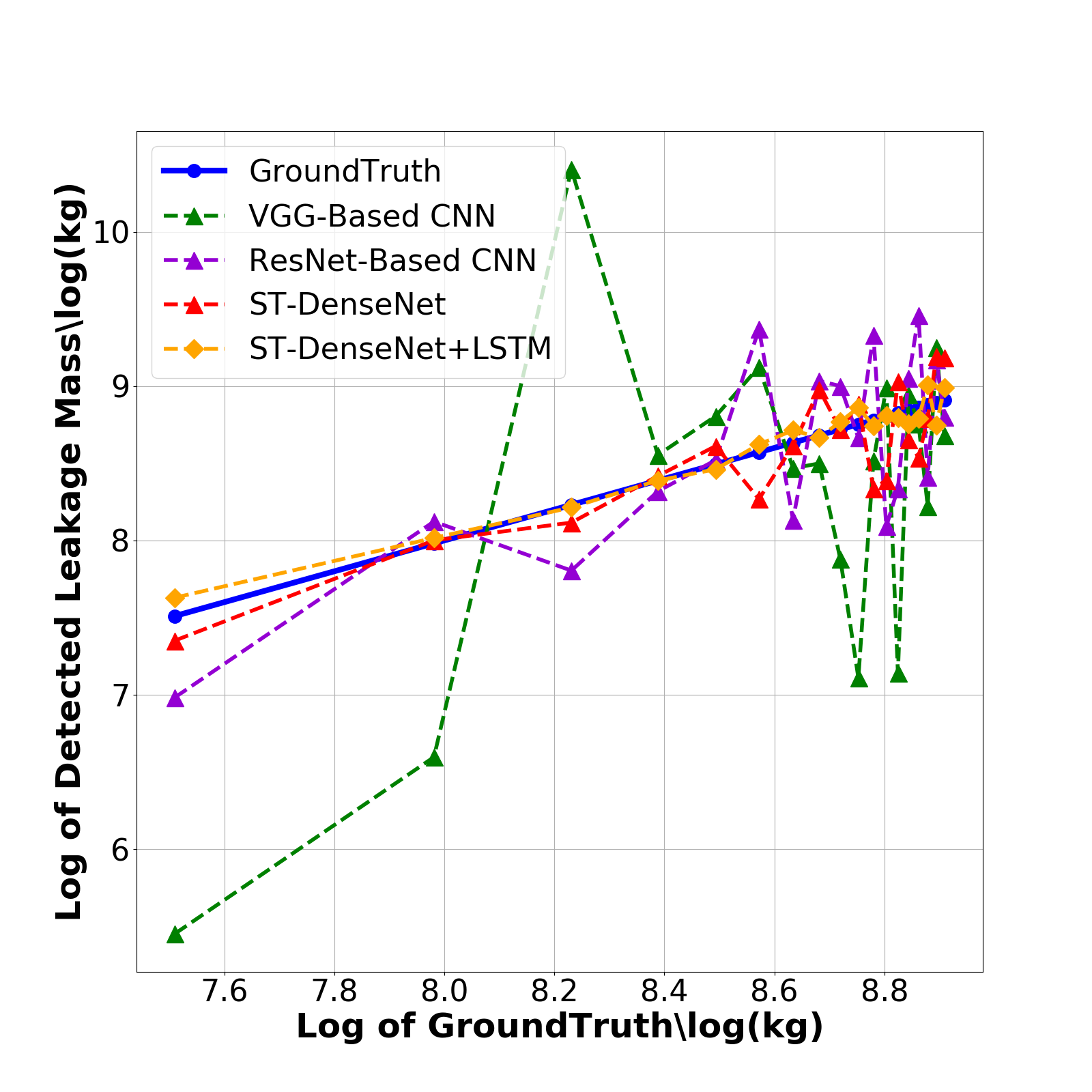}}}
\centerline{
\subfigure[]{\includegraphics[width=0.5\linewidth]{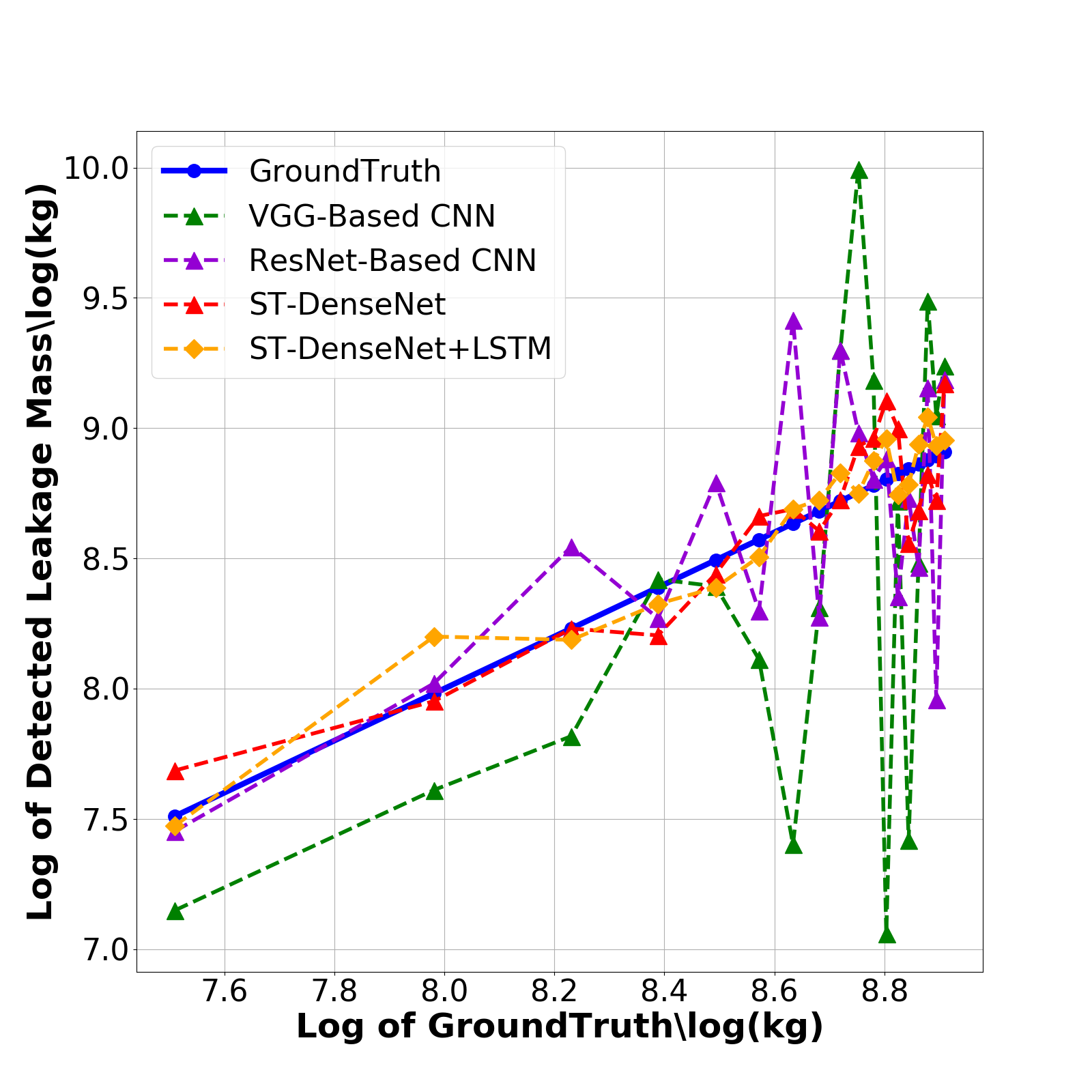}}
\subfigure[]{\includegraphics[width=0.5\linewidth]{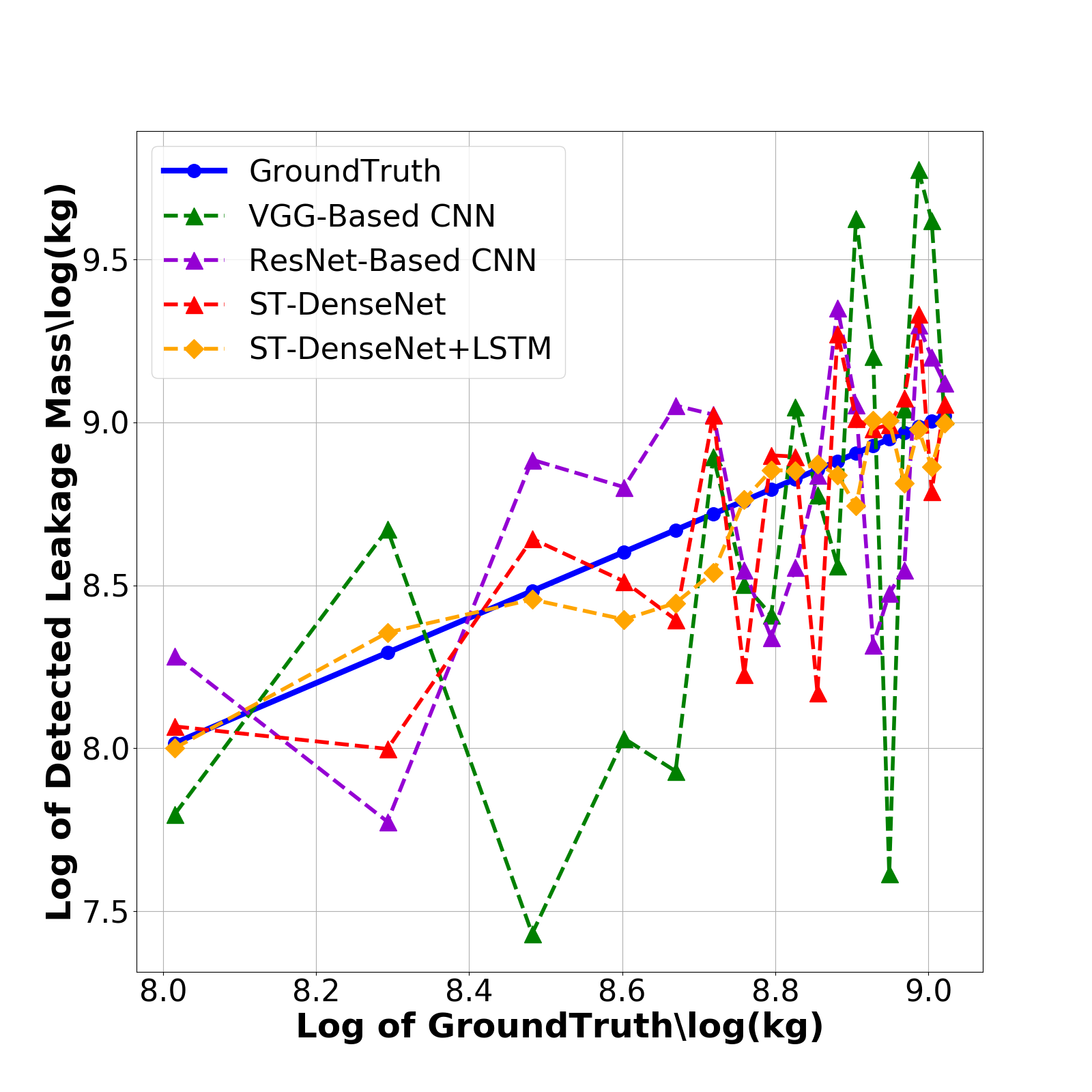}}}
\caption{Illustration of the detection results plotted in the Log of GroundTruth VS. Log of Detected Mass on the data with 30db noise. We provide the enlarged view of the last 19 detection results shown in Fig.~\ref{fig:groundtruth_vs_detected_noise}. The network architectures includes VGGNet~(green), ResNet~(purple), ST-DenseNet~(red), and ST-DenseNet with LSTM~(orange). The ground-truth is in blue. We observe that ST-DenseNet+LSTM model yields the most accurate detection results and the minimum variance among all four methods.}
\label{fig:groundtruth_vs_detected_noise_zoomin}
\end{figure}

We provide the detection results for each method when applied to data with 30db of added noise from Fig.~\ref{fig:noise_detection1} to Fig.~\ref{fig:groundtruth_vs_detected_noise_zoomin}. Specifically, in Fig.~\ref{fig:noise_detection1}, we compare results obtained using ST-DenseNet~(red) to those obtained using VGGNet~(green) and ResNet~(purple) for four different leakage scenarios. We notice that detections results of all three CNN networks including ST-DenseNet become more oscillatory comparing to detection results in Figs.~\ref{fig:exp_detection_mass1}, \ref{fig:exp_detection_mass2}, and  \ref{fig:exp_detection_mass3}. In Figs.~\ref{fig:noise_detection2}, a significant improvement can be observed both in detection accuracy and robustness after ST-DenseNet cascading with LSTM. However, ST-DenseNet still yielded the most accurate results out of all three CNN networks. Similarly, we provide the plots of the results in the Log of GroundTruth VS. Log of Detected Mass of the 200-year leakage process in Fig.~\ref{fig:groundtruth_vs_detected_noise} and the enlarged view of the last 19 detection results in Fig.~\ref{fig:groundtruth_vs_detected_noise_zoomin}. Again, we observe that our ST-DenseNet with LSTM yields the most accurate results compared with all the other detection methods.

\section{Conclusions}
\label{sec:Conclusions}

In this paper, we developed a novel method based on spatial-temporal densely connected convolutional networks (ST-DenseNet) to capture abstract high-level features from the seismic data for the detection of CO$_2$ leakage. To further account for sequential information from CO$_2$ leakage history data, we incorporated long short-term memory networks to ST-DenseNet. Our method not only considers both the spatial and temporal characteristics of seismic data but also significantly reduces the number of parameters in the network's structure. We tested our method using simulated reflection seismic data generated based on various leakage scenarios at the Kimberlina site in the southern San Joaquin Basin, California. By comparing with several commonly-used machine learning methods, we demonstrated that our detection method outperforms traditional regression methods and other popular deep learning models. We also demonstrate the robustness of ST-DenseNet by implementing intra-site cross-location tests and noisy-data tests. Our novel detection method shows great potential in CO$_2$ leakage detection as well as for other monitoring tasks in various subsurface applications. 

\section{ACKNOWLEDGMENTS}

This work was co-funded by the U.S. DOE Office of Fossil Energy’s Carbon Storage program and the Center for Space and Earth Science~(CSES) at Los Alamos National Laboratory (LANL). The computation was performed using super-computers of LANL's Institutional Computing Program.

\appendix
\section{Comparison of Key Parameters at the Three Well Locations - Pressure}
\label{sec:AppendixA}

\begin{table}
\centering
\begin{tabular}{|c|c|c|c|}
\hline
Pressure Change  & Well Location at 1~km & Well Location at 3~km  & Well Location at 6 km \\ 
\hline
\hline
t = 5 yr &  \includegraphics[width=0.2\linewidth, height=0.2\linewidth]{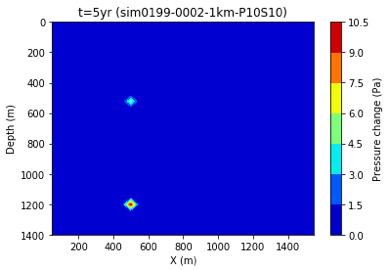} &  \includegraphics[width=0.2\linewidth, height=0.2\linewidth]{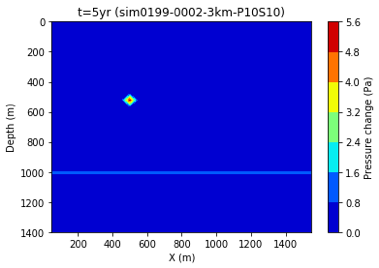}   &  \includegraphics[width=0.2\linewidth, height=0.2\linewidth]{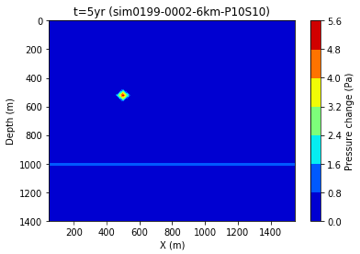} \\
\hline
t = 20 yr & \includegraphics[width=0.2\linewidth, height=0.2\linewidth]{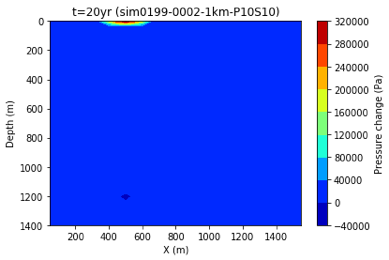} & \includegraphics[width=0.2\linewidth, height=0.2\linewidth]{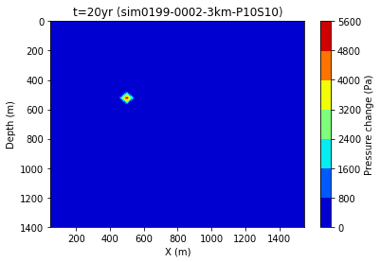} & \includegraphics[width=0.2\linewidth, height=0.2\linewidth]{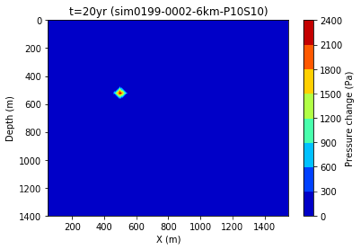}\\
\hline
t = 100 yr & \includegraphics[width=0.2\linewidth, height=0.2\linewidth]{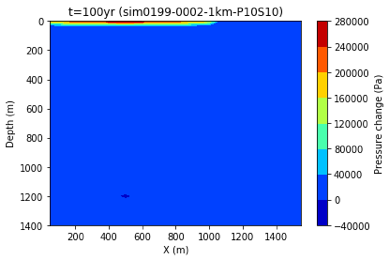} &  \includegraphics[width=0.2\linewidth, height=0.2\linewidth]{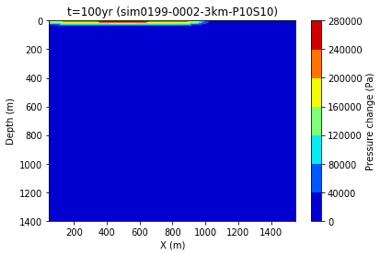} &  \includegraphics[width=0.2\linewidth, height=0.2\linewidth]{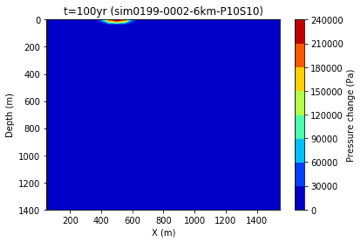}\\
\hline
t = 200 yr & \includegraphics[width=0.2\linewidth, height=0.2\linewidth]{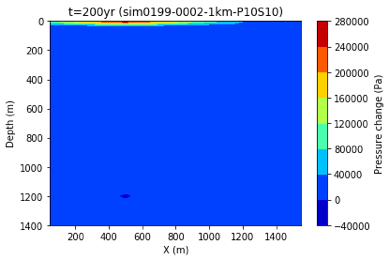}&\includegraphics[width=0.2\linewidth, height=0.2\linewidth]{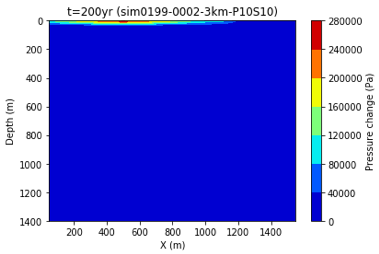} &\includegraphics[width=0.2\linewidth, height=0.2\linewidth]{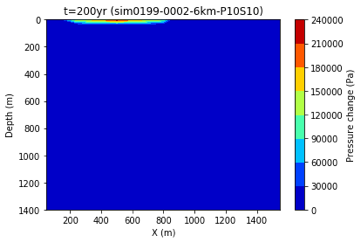}\\
\hline
\end{tabular}
\caption{The illustration of pressure changes at different wells in different year.}
\label{table:ParameterForDifferentWells_Pressure}
\end{table}

\section{Comparison of Key Parameters at the Three Well Locations - $\textrm{CO}_2$ Saturation}
\label{sec:AppendixB}

\begin{table}
\centering
\begin{tabular}{|c|c|c|c|}
\hline
$\mathrm{CO}_2$ Saturation & Well Location at 1~km & Well Location at 3~km  & Well Location at 6 km \\ 
\hline
\hline
t = 5 yr &  \includegraphics[width=0.2\linewidth, height=0.2\linewidth]{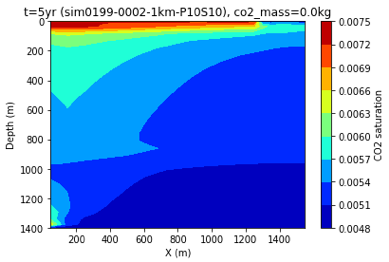} &  \includegraphics[width=0.2\linewidth, height=0.2\linewidth]{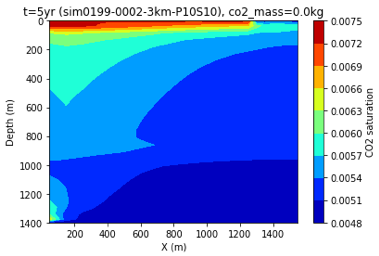}   &  \includegraphics[width=0.2\linewidth, height=0.2\linewidth]{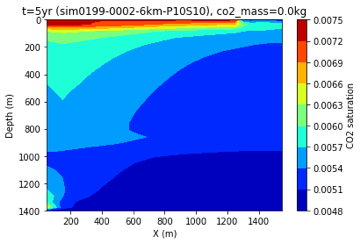} \\
\hline
t = 20 yr & \includegraphics[width=0.2\linewidth, height=0.2\linewidth]{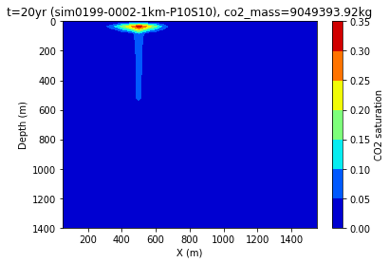} & \includegraphics[width=0.2\linewidth, height=0.2\linewidth]{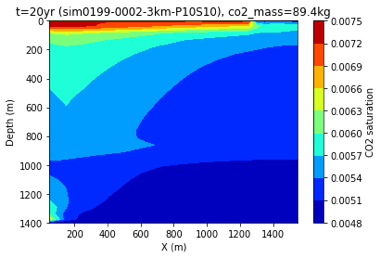} & \includegraphics[width=0.2\linewidth, height=0.2\linewidth]{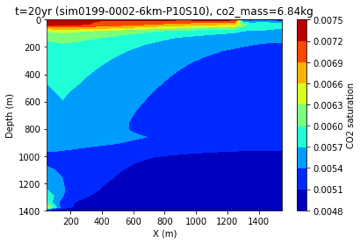}\\
\hline
t = 100 yr & \includegraphics[width=0.2\linewidth, height=0.2\linewidth]{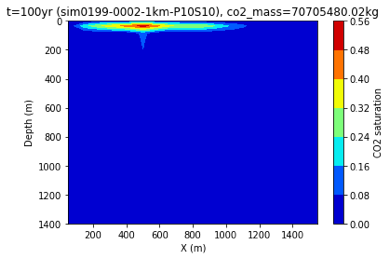} &  \includegraphics[width=0.2\linewidth, height=0.2\linewidth]{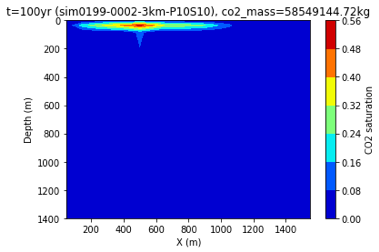} &  \includegraphics[width=0.2\linewidth, height=0.2\linewidth]{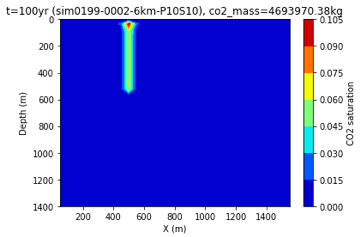}\\
\hline
t = 200 yr & \includegraphics[width=0.2\linewidth, height=0.2\linewidth]{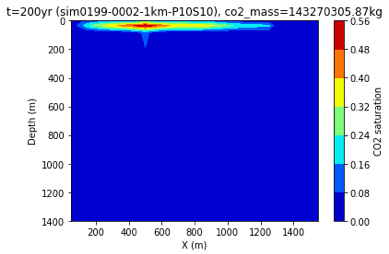}&\includegraphics[width=0.2\linewidth, height=0.2\linewidth]{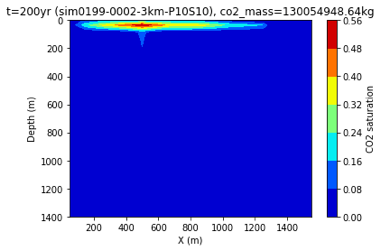} &\includegraphics[width=0.2\linewidth, height=0.2\linewidth]{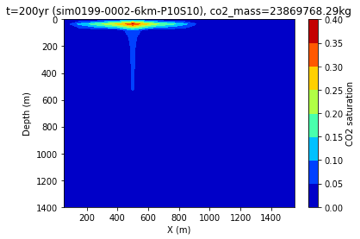}\\
\hline
\end{tabular}
\caption{The illustration of $\mathrm{CO}_2$ Saturation at different wells in different year.}
\label{table:ParameterForDifferentWells_Saturation}
\end{table}

\clearpage

\bibliographystyle{seg}
\bibliography{reference}

\end{document}